\documentclass[aps,prb,twocolumn,showpacs,superscriptaddress,floatfix]{revtex4-1}
\usepackage{dcolumn}
\usepackage{amsmath}
\usepackage{graphicx,epsfig,psfrag}
\usepackage{amssymb}
\usepackage{natbib}
\usepackage{url}
\usepackage[breaklinks=true]{hyperref}
\usepackage{mathtools}
\usepackage{subfigure}
\usepackage{color}
\hypersetup{
        colorlinks=true,
}
\usepackage{bookmark}
\usepackage{epic}

\usepackage{txfonts}
\definecolor{red}{rgb}{1,0,0}
\definecolor{blue}{rgb}{0,0,1}
\definecolor{darkgreen}{rgb}{0,0.5,0}
\definecolor{forestgreen}{rgb}{0.13, 0.55, 0.13}

\begin{document}
\title{Time-dependent spectral functions of the Anderson impurity model in response to a quench with application to time-resolved photoemission spectroscopy}
\author{H. T. M. Nghiem}
\affiliation{
Phenikaa Institute for Advanced Study, Phenikaa University, Yen Nghia, Ha-Dong district, Hanoi 12116, Vietnam}
\affiliation{
Faculty of Basic Science, Phenikaa University, Yen Nghia, Ha-Dong district, Hanoi 12116, Vietnam}
\author{H. T. Dang}
\affiliation{
Phenikaa Institute for Advanced Study, Phenikaa University, Yen Nghia, Ha-Dong district, Hanoi 12116, Vietnam}
\affiliation{
Faculty of Materials Science and Engineering, Phenikaa University, Yen Nghia, Ha-Dong district, Hanoi 12116, Vietnam}
\author{T. A. Costi}
\affiliation
{Peter Gr\"{u}nberg Institut and Institute for Advanced Simulation, 
Research Centre J\"ulich, 52425 J\"ulich, Germany}
\begin{abstract}
We investigate several definitions of the time-dependent spectral function $A(\omega,t)$ of the Anderson impurity model following a quench and within the
  time-dependent numerical renormalization group (TDNRG) method. In terms of the single-particle two-time retarded Green function $G^r(t_1,t_2)$, the definitions we consider
  differ in the choice of the time variable $t$ with respect to $t_1$ and/or $t_2$ (which we refer to as the time reference).  In a previous study  [Nghiem {\it et al.} Phys. Rev. Lett.  {\bf 119}, 156601 (2017)], we investigated the spectral function $A(\omega,t)$, obtained from the Fourier transform of ${\rm Im}[G^r(t_1,t_2)]$ w.r.t. the time difference $t'=t_1-t_2$, with time reference $t=t_2$. Here,
  we complement this work by deriving expressions for the retarded Green function for the choices $t=t_1$ and the average, or Wigner, time $t=(t_1+t_2)/2$, within the TDNRG approach.
  We compare and contrast the resulting $A(\omega,t)$ for the different choices of time reference. While the choice $t=t_1$ results in a spectral function with no time-dependence before the quench ($t<0$) (being identical to the equilibrium initial-state spectral function for $t<0$), the choices $t=(t_1+t_2)/2$ and $t=t_2$ exhibit nontrivial time evolution both before and after the quench.  Expressions for the lesser, greater and advanced Green functions are also derived within TDNRG for all choices of time reference. The average time lesser Green function $G^{<}(\omega,t)$ is particularly interesting, as it
determines the time-dependent occupied density of states $N(\omega,t)=G^{<}(\omega,t)/(2\pi i)$, a quantity that determines the photoemission
current in the context of time-resolved pump-probe photoemission spectroscopy. We present calculations for $N(\omega,t)$ for the Anderson model following a quench, and discuss the resulting time evolution of the spectral features, such as the Kondo resonance and high-energy satellite peaks. We also discuss the issue of thermalization at long times for $N(\omega,t)$.
Finally, we use the results for  $N(\omega,t)$ to calculate the time-resolved photoemission current for the Anderson model following a quench
(acting as the pump) and study the different behaviors that can be observed for different resolution times of a Gaussian probe pulse.
\end{abstract}

\date{\today}

\maketitle
\section{Introduction}

The study of quantum impurity systems out of equilibrium is relevant to several fields, including the nonequilibrium dynamics of ions scattering from metallic surfaces \cite{He2010,Pamperin2015}, the steady state nonequilibrium transport through Kondo-correlated quantum dots \cite{DeFranceschi2002,HanHeary2007,Anders2008a} or the nature of nonequilibrium states in periodically driven quantum dot systems \cite{Schiller2008}.
In addition, solving for the nonequilibrium dynamics of quantum impurity systems is a prerequisite for applications to the nonequilibrium dynamical mean field theory \cite{Freericks2006} of correlated materials, with relevance to interpreting time-resolved photoemission experiments \cite{Perfetti2006,Ligges2018}.
While there are many studies investigating the time-dependent dynamics of quantum impurity systems, including functional and real-time renormalization group methods\cite{Metzner2012,Kennes2012a,Schoeller2009}, flow equation \cite{Lobaskin2005,Wang2010}, quantum Monte Carlo \cite{Gull2011b,Cohen2014a}, density matrix renormalization group methods \cite{Daley2004,White2004,Schmitteckert2010}, hierarchical quantum master equation approach \cite{Schinabeck2016,Schinabeck2018}, and, the time-dependent numerical renormalization group (TDNRG) method \cite{Anders2005,Anders2006,Anders2008a,Anders2008b,Eidelstein2012,Guettge2013,Nghiem2014a,Nghiem2014b,
  Nghiem2016,Nghiem2017,Nghiem2018}, there are fewer studies devoted to investigating the nature of the time-dependent spectral
function in nonequilibrium situations \cite{Nordlander1999,Anders2008b,Cohen2014a,Weymann2015,Bock2016,Nghiem2017,Nghiem2018,Krivenko2019}.

In contrast to the equilibrium case, where the spectral function is uniquely defined via the Lehmann representation, and can be derived directly from the retarded Green function \cite{Bulla2008},  in the case of nonequilibrium, there is a degree of freedom in defining the time-dependent spectral function $A(\omega,t)$ from the Fourier transform of the retarded two-time Green function $G^r(t_1,t_2)$, depending on
how $t$ is measured with respect to $t_1$ and/or $t_2$ prior to carrying out the Fourier transform w.r.t. the relative time $t'=t_1-t_2$.
In the context of time-dependent transport through quantum dots \cite{Jauho1994,Nordlander1999}, the choice $t=t_1$ is appropriate\footnote{\label{Note1}This choice is also appropriate in other situations, e.g., in calculating the injected current from a probe into a Luttinger liquid channel subject to a quench \cite{Calzona2017,Calzona2018,Kennes2014}.}, whereas
in the context of time-resolved photoemission spectroscopy \cite{Freericks2009,Freericks2015,Freericks2017,Randi2017},
the natural choice is the average time $t=(t_1+t_2)/2$.

In this paper, we elaborate more on the various definitions of the time-dependent spectral functions using different time-references, and show the effect of the time-reference on the time evolution of the spectral function of the Anderson impurity model subject to a sudden quench and within the
TDNRG method.  
{We also apply our results of time-dependent spectral function to time-resolved photoemission spectroscopy.}
The outline of the paper is as follows. In Sec.~\ref{sec:model-method-quench-scales}, we describe the model, briefly outline the TDNRG method, define the parameter quench used for all calculations in the paper and specify the relevant time scales.
In Sec. \ref{sec:definitions}, we define the various two-time Green functions studied in this paper, give the various possible definitions of
time-dependent spectral functions $A(\omega,t)$ in terms of the retarded Green function, with time $t$ taken as either $t_1, t_2$ or $(t_1+t_2)/2$, and discuss some general properties. In Sec.~\ref{sec:gf-times} we present expressions for the retarded Green function for each time reference within the TDNRG formalism and discuss their structure and physical interpretation (Sec.~\ref{subsec:expressions}). We show that the average time Green function, like that for $t=t_2$ exhibits a nontrivial time evolution at both negative and positive times (Sec.~\ref{subsec:expressions}).
Numerical issues in the evaluation of time-dependent spectral functions are discussed (Sec.~\ref{subsec:numerical-issues}). In particular, evaluation of the retarded (and also the lesser and greater) Green functions at the average time is  shown to pose a significant numerical bottleneck within TDNRG due to the appearance of 4-loop summations over states which cannot be reduced to
matrix multiplications for efficient evaluation. We resolve this issue by implementing the calculations using parallel computing 
within OpenMP. In Sec.~\ref{subsec:spectra-comparisons}, we evaluate the time-dependent spectral functions numerically for all three time references, for a quench in the Anderson model, and compare the time evolution of the low-energy Kondo resonance and high-energy satellite peaks for the different cases. {In Sec.~\ref{sec:glesser}, we derive expressions for the lesser  Green function at both positive and negative average time (Sec.~\ref{subsec:lgf}) and use these to calculate the time-dependent occupied density of states of the Anderson model following a
quench (Sec.~\ref{subsec:tr-dos}) and the photoemission current intensity for  Gaussian probe-pulses of different widths (Sec.~\ref{subsec:tr-current}).}
Section~\ref{sec:conclusions} concludes with possible future applications of the formalism developed here.
Appendix \ref{sec:appendix-advanced} gives the detailed derivation of the average time advanced Green function, while Appendix \ref{sec:appendix-gf-list} lists the TDNRG expressions for advanced, lesser and greater Green functions for all time references. The convergence of the Lorentzian broadening scheme, used to evaluate the time-dependent spectral functions, is discussed in Appendix \ref{sec:appendix-broadening}, while Appendix \ref{sec:appendix-thermalization} discusses thermalization effects in the time-dependent occupied density of states (lesser Green function) at long times.

\section{Model, method, parameter quench and time scales}
\label{sec:model-method-quench-scales}
\subsection{Model}
\label{subsec:model}
We consider the following time-dependent Anderson impurity model
\begin{align}
  H(t) &= \sum_{\sigma}\varepsilon_d(t)n_{d\sigma}+U(t)n_{d\uparrow}n_{d\downarrow}+\sum_{k\sigma}\epsilon_{k\sigma}c^{\dagger}_{k\sigma}c_{k\sigma}\nonumber\\
       &+\sum_{k\sigma}V(c^{\dagger}_{k\sigma}d_{\sigma}+d^{\dagger}_{\sigma}c_{k\sigma}),\label{eq:siam}
\end{align}
where $\varepsilon_{d}(t)=\theta(-t)\varepsilon_i+\theta(t)\varepsilon_f$ is the energy of the local level, 
$U(t) =\theta(-t)U_i +\theta(t) U_f$ is the local Coulomb interaction, $\sigma$ labels the spin, $n_{d\sigma}=d_{\sigma}^{\dagger}d_{\sigma}$ is the number operator for local electrons with spin $\sigma$,  and $\epsilon_{k}$ is the kinetic energy of the conduction electrons with constant density of states $\rho_{0}(\omega)=\sum_{k}\delta(\omega-\epsilon_{k})=1/(2D)$ with $D=1$ the half-bandwidth. The time-dependence enters via a sudden quench on the model parameters at $t=0$, either by changing the local level position from $\varepsilon_i$ to  $\varepsilon_f$ or  by changing the Coulomb repulsion from $U_i$ to $U_f$ or both. The particular quench studied in this paper is described on more detail at the end of this section.
\subsection{Method}
\label{subsec:method}
We next briefly outline the  TDNRG approach \cite{Anders2005,Anders2006,Nghiem2014a} to the time evolution of physical observables following a
sudden quench at $t=0$. In order to set the notation, we illustrate the approach for a local observable $\hat{O}$. Its time evolution
is given by the expectation value $O(t>0) \equiv \langle\hat{O}\rangle_{\hat{\rho}}= {\rm Tr}\left[e^{-iH^{f}t}\hat{\rho} e^{iH^{f}t}\hat{O}\right]$, where $H_{f}=H(t>0)$ is the final
state Hamiltonian, and $\hat{\rho} = e^{-\beta H_{i}}/Z_i$ is the initial state density matrix corresponding to the initial state Hamiltonian
$H_{i}=H(t<0)$ and $Z_i={\rm Tr}[e^{-\beta H_{i}}]$. Iteratively diagonalizing initial and final state Hamiltonians via the numerical renormalization group (NRG) \cite{KWW1980a,Gonzalez-Buxton1998,Bulla2008} yields the eigenstates and eigenvalues of $H_{i}$ and $H_{f}$ 
on all energy scales and thereby allows $\hat{\rho}$ and the above trace to be calculated. This is accomplished within the complete basis set approach \cite{Anders2005} and yields, within the notation
of Ref.~\onlinecite{Nghiem2014a}, 
\begin{align}
O(t) &=\sum_{m=m_0}^N \sum_{rs\notin KK'}\rho^{i\to f}_{sr}(m) e^{-i(E^m_s-E^m_{r}) t} O^m_{rs} \label{eq:localOt},
\end{align}
in which $m$ labels the iteration, running from the first iteration $m_0$ at which truncation occurs up to a maximum value of $N$, $r$ and $s$ may not both be kept ($K$) states, $O^m_{rs}={_f}\langle lem|\hat{O}|rem\rangle_f$ are the final state matrix elements of $\hat{O}$
at iteration $m$, $E^m_{r}$ are eigenvalues at iteration $m$ and $\rho^{i\to f}_{sr}(m)=\sum_e{_f}\langle sem|\hat{\rho} | rem\rangle_f$ is the initial state density matrix projected onto the final states, with  $\sum_{e}$ denoting the trace over the environment degrees of freedom within the complete basis set
approach \cite{Anders2005}. {Within the latter, the set of discarded states $|lem\rangle$ spans the Hilbert state of all Wilson chains $m=m_0,\dots,N$ diagonalized, resulting in the completeness relation
\begin{align}
&\sum_{m=m_0}^N \sum_{le} | lem\rangle\langle lem|=1,\label{unity-decomposition}
\end{align}
where for $m=N$ all states are counted as discarded (i.e. there are no kept states at iteration $m=N$). 
By using the complete basis set, the initial state density matrix $\hat{\rho}$ appearing in Eq.~(\ref{eq:localOt})
can be represented in terms of shell density matrices $\tilde{\rho}_m$ within the full density matrix approach
\cite{Weichselbaum2007,Peters2006} as
\begin{align}
  \hat{\rho}=\sum_{m=m_0}^N w_m \tilde{\rho}_m,\label{eq:fdm-rho}
\end{align}
with temperature dependent
weights $w_m$ determined via normalization ${\rm Tr [\tilde{\rho}_m]=1}$ (see Refs.~\cite{Weichselbaum2007,Costi2010} for details).}
With the above notation, we proceed in the following sections to the calculation of two-time Green functions within TDNRG which involve calculating expectation values of the form $\langle \hat{O}_{1}(t_1)\hat{O}_2(t_2)\rangle_{\hat{\rho}}$ where $\hat{O}_{1}$ and $\hat{O}_2$ are local operators, e.g., the operators $d_{\sigma}$ and $d^{\dagger}_\sigma$ in (\ref{eq:siam}).
\subsection{Parameter quench}
\label{subsec:quench}
Since the main interest in this paper is to compare the time-dependent spectral functions resulting from different choices of the time reference, we focus on a specific quench on the model (\ref{eq:siam}). We consider switching from a symmetric Kondo regime with $\varepsilon_{i}=-15\Gamma, U_{i}=30\Gamma$ and a vanishingly small Kondo scale $T^{i}_{\rm K}=3\times 10^{-8}D=3\times 10^{-5}\Gamma$ to a symmetric Kondo regime with $\varepsilon_{f}=-6\Gamma$, $U_{f}=12\Gamma$ and a larger Kondo scale $T_{\rm K}=2.5\times 10^{-5}D=2.5\times 10^{-2}\Gamma\gg T^i_{\rm K}= 0.0012 T_{\rm K}$,  and a constant hybridization $\Gamma\equiv\pi\rho_{0}(0)V^2=0.001D$. Thus, the quench is between two
symmetric Kondo states with different degrees of correlation.
\subsection{Time scales}
\label{subsec:time-scales}
The relevant time scales describing the dynamics of the model (\ref{eq:siam}) following the quench specified above are the spin fluctuation time scales $\tau^i_{\rm K}=\hbar/k_{\rm B}T^{i}_{\rm K}$ and $\tau_{\rm K}=\hbar/k_{\rm B}T_{\rm K}$ of the initial and final states, respectively, where $T^{i}_{\rm K}$ and $T_{\rm K}$ are the corresponding initial and final state Kondo temperatures, and the charge fluctuation time scale $\tau_c=\hbar/\Gamma$. The final state Kondo temperature $T_{\rm K}$ is defined via the $T=0$ spin susceptibility $\chi_0$ via $\chi_0=(g\mu_{\rm B})^2/4k_{\rm B}T_{\rm K}$, and similarly with $T^{i}_{\rm K}$.
In the limit of strong correlations $U_{i,f}/\pi\Gamma\gg 1$, the Bethe ansatz expression for $\chi_0$ yields to high accuracy the analytic expression $T_{\rm K}=\sqrt{\Gamma U_f/2}e^{-\pi U_f/8\Gamma + \pi \Gamma/2U_f}$, and a similar expression for $T^i_{\rm K}$ \cite{Zlatic1983,Hewson1997}.
In the following we set all physical constants to unity, i.e.,  $g=\mu_{\rm B}=k_{\rm B}=\hbar = 1$, so expressions such as $tT_{\rm K}$ or $t\Gamma$ should be interpreted, in terms of physical units, as $tk_{\rm B}T_{\rm K}/\hbar$ and $t\Gamma/\hbar$ respectively.

\section{Definitions and general properties}
\label{sec:definitions}
The two-time Green functions of interest to us in this paper,
are the retarded $G^r(t_1,t_2)$, advanced $G^a(t_1,t_2)$, greater $G^>(t_1,t_2)$ and lesser $G^<(t_1,t_2)$ Green functions, which are defined as follows \cite{Jauho2008}
  \begin{align}
    G^r(t_1,t_2)&=-i\theta(t_1-t_2)\langle [d_{\sigma}(t_1),d^{\dagger}_{\sigma}(t_2)]_+\rangle_{\hat{\rho}}\label{eq:gr}\\
    G^a(t_1,t_2)&=+i\theta(t_2-t_1)\langle [d_{\sigma}(t_1),d^{\dagger}_{\sigma}(t_2)]_+\rangle_{\hat{\rho}}\label{eq:ga}\\
    G^>(t_1,t_2)&=-i\langle d_{\sigma}(t_1)d^{\dagger}_{\sigma}(t_2)\rangle_{\hat{\rho}}\label{eq:g>}\\
    G^<(t_1,t_2)&=+i\langle d^{\dagger}_{\sigma}(t_2)d_{\sigma}(t_1)\rangle_{\hat{\rho}}\label{eq:g<}
  \end{align}
  Consider the retarded two-time Green function $G^r(t_1,t_2)=-i\theta(t_1-t_2)\langle [d_{\sigma}(t_1),d^{\dagger}_{\sigma}(t_2)]_+\rangle_{\hat{\rho}}$,
  where the time evolution of the operators may refer to either $H_i$ or $H_f$, depending on whether $t_1,t_2$
  are before or after the quench (which occurs at $t_1=t_2=0$). In the absence of a quench, i.e., in equilibrium $H_i=H_f=H$, we have that 
  $G^r(t_1,t_2)= G^{r}(t_1-t_2,0)=-i\theta(t_1-t_2) \langle [d_{\sigma}(t_1-t_2),d^{\dagger}_{\sigma}(0)]_+\rangle_{\hat{\rho}} $, which depends only on the relative time $t'=t_1-t_2$ and not explicitly on the individual times $t_1$ and $t_2$, and similarly for the other two-time Green functions. Hence, in equilibrium one can define a unique time-independent spectral function $A(\omega)=-\frac{1}{\pi}Im[G^r(\omega+i\eta)]$ with $G^r(\omega+i\eta)\equiv \int_{-\infty}^{+\infty}dt' e^{i(\omega +i\eta)t'}G^{r}(t')$ the Fourier transform of the retarded two-time Green function $G^r(t')\equiv G^r(t',0)$ w.r.t. the relative time $t'$ and $\eta$ is a positive infinitesimal. In contrast, in the presence of a quench, $G^r(t_1,t_2)$ depends explicitly on both $t_1$ and $t_2$, and
  similarly for the other two-time Green functions. Consequently, the Fourier transform $\int_{-\infty}^{+\infty}dt' e^{i(\omega +i\eta)t'}G^{r}(t_1,t_2)$
of $G^r(t_1,t_2)$ w.r.t. $t'=t_1-t_2$ yields a $G^r(\omega,t)$ that can be considered to be a function of either $t=t_1$ (with $t_2=t_1-t'$) or $t=t_2$ (with $t_1=t_2+t'$) or any combination of these $t=t(t_1,t_2)$. The resulting spectral function $A(\omega,t)=-\frac{1}{\pi}Im[G^r(\omega+i\eta,t)]$ then
  has an explicit dependence on the time ``$t$''. The particular choice of $t$ (in terms of $t_1$ and/or $t_2$) results in different spectral functions $A(\omega,t)$, and in this paper
  we shall consider three choices $t=t_1$, $t=t_2$ and $t=(t_1+t_2)/2$. Physically, the different choices describe different 
  processes contributing to the respective spectral functions. Thus, the choice $t=t_1$ would correspond to summing up the amplitudes of all processes in 
  which  a particle is added to the system at some earlier time $t_2<t_1=t$ and then removed at the fixed time $t=t_1$, while the choice $t=t_2$ would 
  correspond to summing up the amplitudes of all processes in which a particle is added at a fixed time $t=t_2$ and is then removed at an arbitrary later time $t_1>t_2$.
  The choice $t=(t_1+t_2)/2$ is the one encountered in time-resolved photoemission spectroscopy \cite{Freericks2009,Randi2017}, see Sec.~\ref{sec:glesser}, while
  the choice  $t=t_1$ is encountered, for example, in time-dependent  transport through quantum dots  \cite{Jauho1994,Nordlander1999}. The choice $t=t_2$ has previously been considered \cite{Anders2008b,Nghiem2017} in the TDNRG to time-evolve spectral functions to infinite times, required, for example, in the
  context of applications to steady state nonequilibrium transport within the scattering states NRG approach \cite{Anders2008a}. Below, we
derive expressions for $A(\omega,t)$ for the choices $t=t_1$ and $t=(t_1+t_2)/2$ within TDNRG and compare these with the results for
the case $t=t_2$ studied in Ref.~\onlinecite{Nghiem2017}.

Before proceeding, we note some general properties. From the definitions (\ref{eq:gr})-(\ref{eq:g<}), we have for all times $t_1,t_2$\cite{Jauho2008}
\begin{align}
  G^r(t_1,t_2) - G^a(t_1,t_2) = G^>(t_1,t_2) - G^<(t_1,t_2), 
\end{align}
and therefore, for any definition of the time, we also have the following after applying the Fourier transform with respect to the time-difference variable
\begin{align}
  G^r(\omega,t) - G^a(\omega,t) = G^>(\omega,t) - G^<(\omega,t).\label{eq:grga}
\end{align}
In cases, where $G^r(\omega,t)=[G^a(\omega,t)]^*$ is satisfied, Eq.~(\ref{eq:grga}) can be used to define the time-dependent
spectral function in terms of the retarded and advanced Green functions, or the lesser and greater Green functions, as
\begin{align}
  A(\omega,t) &= \frac{i}{2\pi}\Big[G^r(\omega,t)-G^a(\omega,t)\Big]\label{eq:Agagr}\\
              &=\frac{i}{2\pi}\Big[G^>(\omega,t) - G^<(\omega,t)\Big],\label{eq:Aglesser}
\end{align}
which are then also equivalent to the definition in terms of the retarded Green function alone,
\begin{align}
A(\omega,t) & =-\frac{Im[G^r(\omega,t)]}{\pi}.\label{eq:Agr}
\end{align}
The condition $G^r(\omega,t)=[G^a(\omega,t)]^*$ is satisfied for
the case  $t=(t_1+t_2)/2$. To see this, we consider the retarded and advanced Green functions in terms of the relative ($t'=t_1-t_2$) and average time $t$, i.e., $G^r(t',t)=-i\theta(t') \langle [d_{\sigma}(t+t'/2),d^{\dagger}_{\sigma}(t-t'/2)]_+\rangle_{\hat{\rho}}$
and $G^a(t',t)=+i\theta(-t') \langle [d_{\sigma}(t+t'/2),d^{\dagger}_{\sigma}(t-t'/2)]_+\rangle_{\hat{\rho}}$. It then follows that $[G^a(t',t)]^*=G^r(-t',t)$, which upon Fourier transforming w.r.t. $t'$ gives $G^r(\omega,t)=[G^a(\omega,t)]^*$. This allows a unique real spectral function to be defined for arbitrary time $t$ using either Eqs.~(\ref{eq:Agagr})-(\ref{eq:Aglesser}) or Eq. ~(\ref{eq:Agr}).
In contrast, one cannot define the spectral function  using Eqs.~(\ref{eq:Agagr})-(\ref{eq:Aglesser}) for the cases with time set to either $t_1$ or $t_2$ since then $G^r(\omega,t)=[G^a(\omega,t)]^*$
is not guaranteed to hold for all times $t$. In these cases, the time-dependent spectral function is defined as in Eq.~(\ref{eq:Agr}) in terms of the imaginary part of the retarded Green function,
i.e., $A(\omega,t) =-{{\rm Im}[G^r(\omega,t)]}/{\pi}$.

In equilibrium, $H_i=H_f$, $G^<(\omega)$ and $G^>(\omega)$ are related to the equilibrium spectral
function $A(\omega)$ via
\begin{align}
  G^<(\omega) & = 2\pi i f(\omega) A(\omega),\label{eq:fd1}\\
  G^>(\omega) & = -2\pi i [1-f(\omega)] A(\omega),\label{eq:fd2}
  \end{align}
where $f(\omega)$ is the Fermi function. Eqs.~(\ref{eq:fd1})-(\ref{eq:fd2})
reflect the fluctuation-dissipation theorem relating correlation functions [$G^<$ and $G^>$] to dissipation [$A(\omega)\propto {\rm Im}[G^r(\omega)$]\cite{Jauho2008}. In nonequilibrium, these relations no longer hold for arbitrary times. Consider for example, Eq.~(\ref{eq:fd1}).
Looking at the expression for $G^<(\omega,t)$ and $A(\omega,t)$ at the average time $t=(t_1+t_2)/2 =0$ within an arbitrary complete basis set of eigenstates $|m\rangle_i$ of $H_{i}$ with eigenvalue $E_m$ and an arbitrary complete set of eigenstates $|m_1\rangle_f$ of $H_f$ with eigenvalues $E_{m_1}$, we find
\begin{align}
A(\omega,t=0)=&\sum_{mnm_1n_1} {_i}\langle m|m_1\rangle_f B_{m_1n_1}{_f}\langle n_1|n\rangle_i C_{nm}\frac{(e^{-\beta E_m}+e^{-\beta E_n})}{Z_i}\nonumber\\
&\times \delta(\omega-\frac{E_{n_1}-E_{m_1}}{2}-\frac{E_{n}-E_{m}}{2}),\label{eq:Azero}\\
G^<(\omega,t=0)=&2\pi i\sum_{mnm_1n_1} {_i}\langle m|m_1\rangle_f B_{m_1n_1}{_f}\langle n_1|n\rangle_i C_{nm}\frac{e^{-\beta E_n}}{Z_i}\nonumber\\
&\times \delta(\omega-\frac{E_{n_1}-E_{m_1}}{2}-\frac{E_{n}-E_{m}}{2}).\label{eq:Gzero}
\end{align}
In the above, and throughout this paper, we set $B=d_{\sigma}$ and $C=d_{\sigma}^{\dagger}$, with matrix elements
denoted by $B_{m_1n_1}=\langle m_1|B|n_1\rangle$ and $C_{mn}=\langle m|C|n\rangle$. From (\ref{eq:Azero})-(\ref{eq:Gzero}), one can directly verify that $G^<(\omega,t=0)=2\pi i f(\omega)A(\omega,t=0)$ is only satisfied when $|m\rangle_i=|m_1\rangle_f$ and $E_m=E_{m_1}$, which is equivalent to the equilibrium case $H_i=H_f$. This shows that the fluctuation-dissipation theorem as expressed in Eq.~(\ref{eq:fd1}) is not valid in nonequilibrium. We return to Eq.~(\ref{eq:fd1}) in Sec.~\ref{sec:glesser} and in Appendix~\ref{sec:appendix-thermalization} when
we discuss thermalization in the long-time limit.

\section{Retarded Green functions and spectral functions for different time references}
\label{sec:gf-times}
In this section we give the TDNRG expressions for the retarded Green function $G^r(\omega+i\eta,t)$
for the three time references $t=t_1$, $t=(t_1+t_2)/2$ and $t=t_2$, at both positive and negative times, and
interpret the different expressions physically (Sec.~\ref{subsec:expressions}).
The derivation for the case $t=t_2$ has been given in detail elsewhere\footnote{\label{Note2}See Supplementary Material of Ref.~\onlinecite{Nghiem2017} for the detailed derivation of the retarded Green function with  time measured relative to $t_2$ and for the spectral weight sum rule.}, and the derivations for the other cases
  are similar, e.g., the derivation of the average time retarded Green function can be carried out following the detailed derivation of the corresponding advanced Green function in Appendix~\ref{sec:appendix-advanced}.  Appendix~\ref{sec:appendix-gf-list} lists the TDNRG expressions for the advanced, lesser and greater Green functions for all three time references.
Numerical issues in the evaluation of the resulting time-dependent spectral functions are discussed in Sec.~\ref{subsec:numerical-issues}. 
Finally, Sec.~\ref{subsec:spectra-comparisons} compares the numerical results for the spectral functions $A(\omega,t)$ for the three time references.

\subsection{Retarded Green function expressions}
\label{subsec:expressions}
For positive times $t=t_1>0$, we have in  the notation of Refs.~\onlinecite{Nghiem2014a,Nghiem2017} (see
  also Refs.~\onlinecite{Anders2008b,Weymann2015})
  \begin{align}
&G^r(\omega+i\eta,t=t_1>0) 
                =\sum_{m}\Big\{\sum_{rsq}^{\notin KK'K''}\Big[B^m_{rs}\rho_{sq}^{i\to f}(m)e^{i(E_{q}^{m}-E_{s}^{m})t}\nonumber\\
                +& \rho_{rs}^{i\to f}(m)e^{i(E_{s}^{m}-E_{r}^{m})t}B^m_{sq}\Big]\frac{C^m_{qr}}{\omega-E^m_q+E^m_r+i\eta}(1-e^{i(\omega-E^m_q+E^m_r+i\eta)t})\nonumber\\
+&\sum_{rsr_1s_1}^{\notin KK'K_1K'_1} S^m_{rr_1}B^m_{r_1s_1}e^{i(\omega+E^m_{r_1}-E^m_{s_1}+i\eta)t}S^m_{s_1s}\frac{\sum_{q}(C^m_{sq}\tilde{R}^m_{qr}+\tilde{R}^m_{sq}C^m_{qr})}{\omega-E^m_{s}+E^m_{r}+i\eta}\Big\},\label{eq:gf-positive-t1}
\end{align}
to be compared with the analogous expression at $t=(t_1+t_2)/2>0$, 
\begin{align}
  G^r&(\omega,t=(t_1+t_2)/2>0)=\nonumber\\
  &\sum_{m}\sum_{rsq}^{\notin KK'K''}  {\rho}^{i\to f}_{rs}(m) 
\Big[\frac{B^m_{sq} C^m_{qr}(e^{i(E^m_s-E^m_r)t} - e^{2i[\omega+E^m_s-E^m_q+i\eta]t} )}{\omega+(E^m_s+E^m_r)/2-E^m_q+i\eta}\nonumber\\
+&\frac{C^m_{sq}  B^m_{qr}(e^{i(E^m_s-E^m_r)t} -e^{2i[\omega-E^m_r+E^m_q+i\eta]t})}{\omega-(E^m_s+E^m_r)/2+E^m_q+i\eta}\Big]\nonumber\\
  +&\sum_{m}\sum_{rsr_1s_1}^{\notin KK'K_1K_1'}\frac{S^m_{r_1r}B^m_{rs}S^m_{ss_1} \sum_qC^m_{s_1q}\tilde{R}^m_{qr_1} e^{2i(\omega+(E^m_r-E^m_s)+i\eta)t}}{\omega+(E^m_r-E^m_s)/2-(E^m_{s_1}-E^m_{r_1})/2+i\eta}\nonumber\\
  +&\sum_{m}\sum_{rsr_1s_1}^{\notin KK'K_1K_1'}\frac{S^m_{r_1r}B^m_{rs}S^m_{ss_1} \sum_q\tilde{R}^m_{s_1q}C^m_{qr_1} e^{2i(\omega+(E^m_r-E^m_s)+i\eta)t}}{\omega+(E^m_r-E^m_s)/2-(E^m_{s_1}-E^m_{r_1})/2+i\eta}, \label{eq:gf-positive-t12}
\end{align}
and the expression for the case $t=t_2>0$ \cite{Nghiem2017}, 
\begin{align}
&G^r(\omega+i\eta,t=t_2>0) 
=\sum_{m}\sum_{rsq}^{\notin KK'K''}\rho_{sr}^{i\to f}(m)e^{-i(E_{s}^{m}-E_{r}^{m})t}\nonumber\\
&\times\Big(\frac{B^m_{rq}C^m_{qs}}{\omega+E^m_r-E^m_q+i\eta}
+\frac{C^m_{rq} B^m_{qs}}{\omega+E^m_q-E^m_s+i\eta}\Big),\label{eq:gf-positive-t2}
\end{align} 
In the above, $\rho_{sq}^{i\to f}(m)$ is the full reduced density matrix of the initial state projected onto the final states (here $|s\rangle$ and $|q\rangle$), $\tilde{R}^m_{qr}$ is the full reduced density matrix of the initial state\cite{Nghiem2014a} (i.e., $q,r$ label initial states), and $S^m_{rr_1}$ are overlap matrix elements between initial and final states (whether $r$ or $r_1$ is the initial state can be deduced by examining how the indices appear in $\tilde{R}$ or $\rho^{i\to f}$).

All three expressions (\ref{eq:gf-positive-t1})-(\ref{eq:gf-positive-t2}) yield the same final state Green function in the infinite time limit (noting that many terms decay to zero as $e^{-\eta t}$),
\begin{align}
&G^r(\omega+i\eta,t=+\infty) 
                =\sum_{m}\sum_{rs}^{\notin KK'}\frac{B^m_{rs}C^m_{sr}[\rho_{ss}^{i\to f}(m)+\rho_{rr}^{i\to f}(m)]}{\omega+E^m_r-E^m_s+i\eta},\label{eq:gf-positive-inf}
\end{align}
and hence also the same final state spectral function $A(\omega,t=+\infty)=-{\rm Im}[G^r(\omega,t=+\infty)]/\pi$ in this limit.

At finite times, we can interpret the expressions  (\ref{eq:gf-positive-t1})-(\ref{eq:gf-positive-t2}) physically as follows. Starting with
Eq.~(\ref{eq:gf-positive-t1}) for $t=t_1>0$ we note that the first term in square brackets, involving final state excitations at $\omega = E^m_q-E^m_r$, describes, with increasing time, the evolution towards the final state at $t=+\infty$ resulting in Eq.~(\ref{eq:gf-positive-inf}), while the last term, containing initial state excitations at $\omega = E^m_s-E^m_r$ and weighted by the factor $e^{-\eta t}$, describes the decay of initial state contributions with increasing time.
Similarly, for the average time Green function in Eq,~(\ref{eq:gf-positive-t12}) we see that the first term in square brackets, involving final state excitations at $\omega = \pm [(E^m_s+E^m_r)/2 -E^m_q]$, describes the evolution towards the final state and results in Eq.~(\ref{eq:gf-positive-inf}) at $t=+\infty$, while the last terms, involving a sum of initial and final state excitations at $\omega=-[(E^m_r-E^m_s)/2-(E^m_{s_1}-E^m_{r_1})/2]$ and weighted by the factor $e^{-2\eta t}$ describe the decay of initial state contributions with increasing time. Finally, the single term in the Green function for $t=t_2$ in Eq.~(\ref{eq:gf-positive-t2}), containing only final state excitations, is seen to describe the evolution towards the final state at $t=+\infty$ described
by Eq.~(\ref{eq:gf-positive-inf}) . Since both times are always positive (i.e., $t_1>t_2=t>0$) in arriving at  Eq.~(\ref{eq:gf-positive-t2}),
the influence of the initial state on the positive time evolution is entirely contained in the projected density
matrix $\rho_{sr}^{i\to f}(m)$.

We next consider the negative time expressions for the retarded Green functions for the different time references.
For $t=t_1<0$, we notice both times are always negative ($0>t=t_1>t_2$), and hence the dynamics of the
operators $d_{\sigma}(t_1)$ and $d_{\sigma}^{\dagger}(t_2)$, appearing in the definition of $G^r(t_1,t_2)$, 
is governed solely by the initial state Hamiltonian. Therefore, the expression for the spectral function at $t<0$ for $t=t_1$ is identical to the equilibrium initial state spectral function, which has no $t$-dependence and is given by
\begin{align}
&G^r(\omega,t=t_1 <0)=
                \sum_{m}\sum_{rs}^{\notin KK'}\frac{B^m_{rs}\sum_q (C^m_{sq}\tilde{R}^m_{qr}+\tilde{R}^m_{sq}C^m_{qr})}{\omega+E^m_r-E^m_s+i\eta}.\label{eq:gf-negative-t1}
\end{align}
In contrast, the analogous expressions for the cases $t=(t_1+t_2)/2 <0$ and $t=t_2<0$ show a non-trivial dynamics also 
at negative times. For $t=(t_1+t_2)/2 <0$, we have 
\begin{align}
&G^r(\omega,t=(t_1+t_2)/2<0)=\nonumber\\
&\sum_{m}\sum_{rs}^{\notin KK'}\frac{B^m_{rs}\sum_q (C^m_{sq}\tilde{R}^m_{qr}+\tilde{R}^m_{sq}C^m_{qr})}{\omega+E^m_r-E^m_s+i\eta} (1-e^{-2i(\omega+E^m_r-E^m_s+i\eta)t})\nonumber\\
                      +&\sum_{m}\sum_{rsr_1s_1}^{\notin KK'K_1K_1'}\frac{S^m_{rr_1}B^m_{r_1s_1}S^m_{s_1s} \sum_q(C^m_{sq}\tilde{R}^m_{qr}+\tilde{R}^m_{sq}C^m_{qr}) e^{-2i(\omega-(E^m_{s}-E^m_{r})+i\eta)t} }{\omega+(E^m_{r_1}-E^m_{s_1})/2-(E^m_{s}-E^m_{r})/2+i\eta} \label{eq:gf-negative-t12},  
\end{align}
while the expression for $t=t_2<0$ has been derived in Ref.~\onlinecite{Nghiem2017} and is given by 
\begin{align}
&G^r(\omega,t=t_2<0)=\sum_m\Big[\sum_{rs}^{\notin KK'}\frac{B^m_{rs}(1-e^{-i(\omega+E^m_r-E^m_s+i\eta)t})}{\omega+E^m_r-E^m_s+i\eta}\nonumber\\
  &+\sum_{rsr_1s_1}^{\notin KK'K_1K'_1} \frac{S^m_{rr_1}B^m_{r_1s_1}S^m_{s_1s}e^{-i(\omega+E^m_r-E^m_s+i\eta)t}}{\omega+E^m_{r_1}-E^m_{s_1}+i\eta}\Big]\nonumber\\
  &\times\sum_{q}(C^m_{sq}\tilde{R}^m_{qr}+\tilde{R}^m_{sq}C^m_{qr}).\label{eq:gf-negative-t2}
\end{align}
All three expressions (\ref{eq:gf-negative-t1})-(\ref{eq:gf-negative-t2}) reduce to the initial state Green function at $t=-\infty$
which is given by the time-independent expression in Eq.~(\ref{eq:gf-negative-t1}), i.e.,
\begin{align}
&G^r(\omega,t=-\infty)=
                \sum_{m}\sum_{rs}^{\notin KK'}\frac{B^m_{rs}\sum_q (C^m_{sq}\tilde{R}^m_{qr}+\tilde{R}^m_{sq}C^m_{qr})}{\omega+E^m_r-E^m_s+i\eta}. \label{eq:gf-negative-inf}
\end{align}

The structure of Eqs.~(\ref{eq:gf-negative-t12}) and (\ref{eq:gf-negative-t2}) can be interpreted as follows. The first terms
in both expressions involving initial state excitations at $\omega=E^m_s-E^m_r$ 
describe the evolution towards the initial state Green function in Eq.~(\ref{eq:gf-negative-t1}) as $t\to -\infty$,
while the second terms in these expressions involving mixed initial and final state excitations at  $\omega=-[(E^m_{r_1}-E^m_{s_1})/2-(E^m_{s}-E^m_{r})/2]$ (weighted by $e^{2\eta t}$) and final state excitations at $\omega=(E^m_{s_1}-E^m_{r_1})$ (weighted by $e^{\eta t}$) describe the decay of final state
contributions with increasing negative time. We note also that the negative time Green function at $t=t_2$ in Eq.~(\ref{eq:gf-negative-t2}) resembles the positive time Green function for $t=t_1$  in 
Eq.~(\ref{eq:gf-positive-t1}). Finally, as for the Green function at $t=t_2$ \cite{Nghiem2017}, one can
show that the retarded Green functions for the other time references satisfy the spectral weight sum rule 
  \begin{align}
  -\int_{-\infty}^{+\infty}\frac{{\rm Im}[G^r(\omega,t)]}{\pi}d\omega &= 1\label{eq:sumrule}.
  \end{align}

  \subsection{Numerical issues}
  \label{subsec:numerical-issues}
Before presenting the numerical results for the spectral functions at different time references, we first address two numerical issues that arise in these calculations. First, in the numerical evaluations of the time-dependent spectral functions (Sec.~\ref{subsec:spectra-comparisons}) a broadening
procedure has to be applied to the imaginary parts of the expressions (\ref{eq:gf-positive-t1})-(\ref{eq:gf-positive-t2}) and (\ref{eq:gf-negative-t1})-(\ref{eq:gf-negative-t2}) in order to obtain smooth spectral functions $A(\omega,t)$ from the discrete representations of the Green functions. While the usual Gaussian or logarithmic-Gaussian schemes \cite{Sakai1989,Bulla2001,Bulla2008} can be applied to (\ref{eq:gf-positive-t2})  and (\ref{eq:gf-negative-t1}), which have the usual pole structure, a different procedure is required for the other expressions (\ref{eq:gf-positive-t1})-(\ref{eq:gf-positive-t12}) and (\ref{eq:gf-negative-t12})-(\ref{eq:gf-negative-t2}).
The reason is that the latter expressions contribute to the imaginary part of $G^r(\omega,t)$ both a regular part (from the first terms in these expressions) and also a set of delta functions (from the poles in the second terms), see also the discussion in Ref.~\onlinecite{Note2}. In addition, the infinitesimal $\eta$ in the expressions (\ref{eq:gf-positive-t1})-(\ref{eq:gf-positive-t12}) and (\ref{eq:gf-negative-t12})-(\ref{eq:gf-negative-t2})
  occurs also in the time evolution factors in the numerators. Its presence there is important in capturing the growth/decay of final/initial state contributions as discussed in detail above. A consistent scheme to evaluate both the regular and pole contributions to the expressions (\ref{eq:gf-positive-t1})-(\ref{eq:gf-positive-t12}) and (\ref{eq:gf-negative-t12})-(\ref{eq:gf-negative-t2}) is to set
  $\eta$ to a small finite value throughout. For the pole contribution, this approach would correspond to a Lorentzian broadening scheme, so we shall henceforth
  denote this scheme as the Lorentzian broadening approach to time-dependent spectral functions. Specifically, 
  we set $\eta=\eta_0|\Delta E|$ where $\Delta E=E_p^m -E_q^m$ is an excitation energy and $\eta_0$ is the broadening parameter,
which is usually taken as $\eta_0=1/N_z$ where $N_z$ is the number of values used in the $z$-averaging procedure \cite{Oliveira1994,Campo2005}. 
For the remaining expressions (\ref{eq:gf-positive-t2}) and (\ref{eq:gf-negative-t1}), the usual logarithmic-Gaussian broadening procedure can be applied, $$\delta(\omega-\Delta E)\to \frac{e^{-\eta_0^2/4}}{\eta_0|\Delta E|\sqrt{\pi}} e^{-[\ln (|\omega/\Delta E|)/\eta_0]^2}.$$
The dependence of the results on $\eta_0$ within the logarithmic-Gaussian broadening is weak and values as large as $\eta_0=0.3$ suffice for convergence (see Fig.~\ref{fig:fig12} of Appendix~\ref{sec:appendix-broadening})\footnote{\label{Note3}For the logarithmic-Gaussian broadening, the notation $b=\eta_0$ is also encountered in the literature.}. For the Lorentzian broadening scheme, the dependence of the results on $\eta_0$ is stronger and convergence w.r.t. $\eta_0$ needs to be
checked explicitly. From Appendix~\ref{sec:appendix-broadening}, we show that converged results are obtained by choosing $\eta_0=1/N_z$ with $N_z\geq 32$, i.e., a broadening parameter $\eta_0=0.03125$ suffices for converged results within the Lorentzian broadening scheme.

A second issue in the evaluation of the average time expressions (\ref{eq:gf-positive-t12}) and (\ref{eq:gf-negative-t12}) as well as the expressions for the lesser Green functions (\ref{eq:glesser-positive-t12})-(\ref{eq:glesser-negative-t12}) in Sec.~\ref{sec:glesser} is the significant numerical challenge in evaluating these expressions as compared to the evaluation of the Green functions with time reference $t=t_1$ or $t=t_2$. This is due to the summations over four different indices in the former expressions, in which the appearance of all the four indices in the denominators of these expressions, prohibits recasting these summations as matrix multiplications for efficient evaluation within the optimized Basic Linear Algebra Subprograms (BLAS) package. For a given calculation, the time consumption in calculating the terms with four loops of the above kind is estimated to be about $100\sim 200$ times longer than calculating the terms with three loops. 
In order overcome this computational bottleneck and to make the calculation of the 4-loop terms feasible, we use OpenMP parallelization, in which the total sum is divided into smaller tasks calculated in individual threads [e.g., OpenMP applied to the loop $\sum_{r}$ in the last two terms of  Eqs.~(\ref{eq:gf-positive-t12}) and (\ref{eq:gf-negative-t12})]. These threads utilize common data and process the different tasks in the resulting partial sums independently and hence there is no overhead from communication between the threads. Therefore, the time consumption decreases linearly with increasing  number of threads used in the paralleling computation and makes the calculation of the average (and lesser) Green functions feasible.

\subsection{Comparison of Spectral functions for different time references}
\label{subsec:spectra-comparisons}
\begin{figure}[t]
  \includegraphics[width=0.45\textwidth]{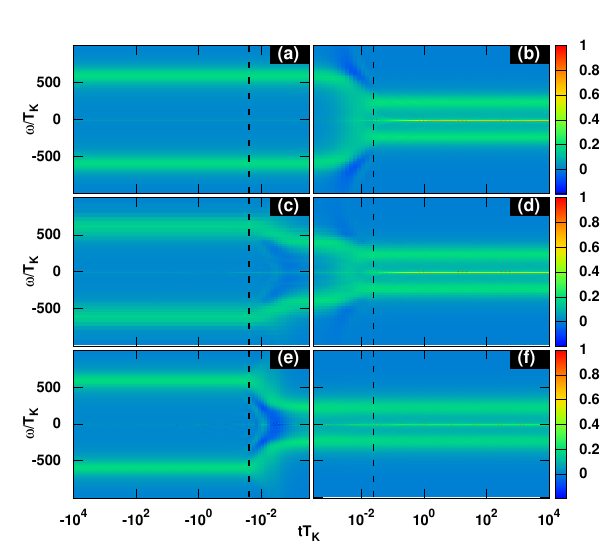}
  \caption 
    {Time evolution of the normalized spectral function $\pi\Gamma A(\omega,t)$ vs $tT_{\rm K}$ from negative times (left panels), to positive times (right panels) 
    for the symmetric Anderson model subject to a quench at $t=0$ specified by  $\varepsilon_{i}=-15\Gamma,U_{i}=30\Gamma$ and 
      $\varepsilon_{f}=-6\Gamma,U_{f}=12\Gamma$ with final state Kondo temperature $T_{\rm K}=2.5 \times 10^{-5}D=2.5\times 10^{-2}\Gamma$, and on a linear frequency scale. Top panels (a) and (b) use as time reference $t=t_1$, middle panels (c) and (d) use $t=(t_1+t_2)/2$ and lower panels (e) and (f) use $t=t_2$. 
Dashed lines mark 
      $t\Gamma=\pm 1$ ($tT_{\rm K}=\pm 10^{-1}$). The spectral function at negative time is time-independent for $t=t_1$ (top panels) and 
      time-dependent for $t=(t_1+t_2)/2$ (middle panels) and $t=t_2$ (lower panels). The high-energy satellite peaks shift from their initial state 
    ($\omega=\pm \varepsilon_{i}=\pm 15\Gamma\approx \pm 600T_{\rm K}$) to their final state values 
      ($\omega=\pm\varepsilon_{f}=\pm 6\Gamma\approx \pm 240 T_{\rm K}$) in the positive time range $10^{-3}/T_{\rm K}<t<2.5\times 10^{-2}/T_{\rm K}=1/\Gamma$ for $t=t_1$ (top panels), in a similar negative time range for $t=t_2$ (lower panels) and in both the above time ranges for $t=(t_1+t_2)/2$ (middle panels). The TDNRG calculations use a discretization parameter $\Lambda=4$, $z$ averaging \cite{Oliveira1994,Campo2005} with $N_z=32$ and a cutoff energy $E_{cut}=24$. {Results for $t=t_2$ in (e) and (f) are from Ref.~\onlinecite{Nghiem2017} and are included here for the purpose of comparison.}} 
    \label{fig:fig1}
  \end{figure}
  \begin{figure}[t]
  \includegraphics[width=0.45\textwidth]{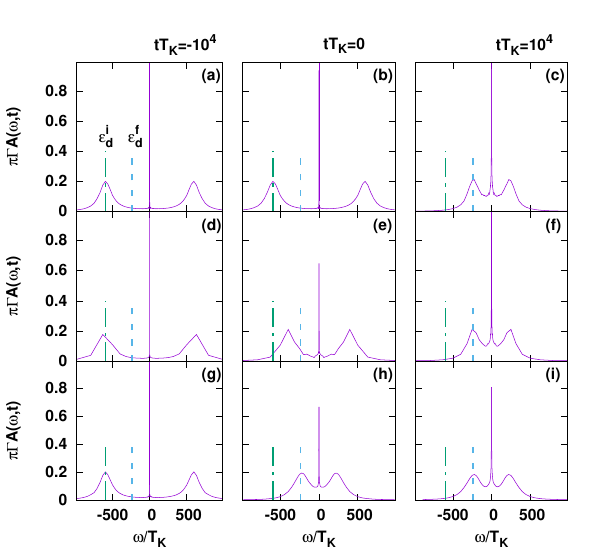}
  \caption 
    {Cuts of the spectral function in Fig.~\ref{fig:fig1} at selected (fixed) times $tT_{\rm K}=-10^{4}$ (left panels), $tT_{\rm K}=0$ (middle panels) and  $tT_{\rm K}=+10^{4}$ (right panels). Top panels refer to $t=t_1$, middle panels to $t=(t_1+t_2)/2$ and lower panels to   $t=t_2$ \cite{Nghiem2017}. Initial, $\varepsilon_{i}=-15\Gamma\approx -600T_{\rm K}$, and final, $\varepsilon_{f}=-6\Gamma\approx -240T_{\rm K}$, state positions of the local level are indicated with vertical dot-dashed and dashed lines respectively. {Results for $t=t_2$ in (g)-(i) are from Ref.~\onlinecite{Nghiem2017} and are included here for the purpose of comparison.}}  
    \label{fig:fig2}
\end{figure} 
{It is instructive to compare the new results of this paper for the spectral functions  $A(\omega,t) =-{\rm Im}[G^r(\omega,t)]/\pi$ at finite times $t=t_1$ and $t=(t_1+t_2)/2$ with our previous results for the same quantity and for the same quench on the Anderson model described in Sec.~\ref{sec:model-method-quench-scales}, but calculated at $t=t_2$ \cite{Nghiem2017}. Figure~\ref{fig:fig1}  compares the overall time-evolution of these spectral functions at the three time references $t=t_1$ (top panels), at $t=(t_1+t_2)/2$ (middle panels) and at $t=t_2$ (lower panels)\cite{Nghiem2017}}.
All cases exhibit both high-energy features (satellite peaks) and a low-energy feature around the Fermi level, the Kondo resonance (to which we shall return to below in more detail).  
The presence of time evolution at negative times for the cases of average time [Figure~\ref{fig:fig1}(c)]and  for $t=t_2$ [Fig.~\ref{fig:fig1}(e)] 
and its absence for the case $t=t_1$ [Fig.~\ref{fig:fig1}(a)] is clearly visible. The nontrivial  dynamics at negative times for  the former cases does not violate causality. It simply reflects the fact that upon Fourier transforming $G^r(t_1,t_2)$ w.r.t. $t'=t_1-t_2>0$ to obtain $A(\omega,t)$ one picks up contributions from both initial states (when $0>t_1>t_2=t$) and final states (when $t_1>0>t_2=t$). A common feature of all three spectral functions is that
the largest rearrangement of spectral weight, which is associated with a shift of the satellite peaks from
$\omega=\varepsilon^i_d$ (and $\varepsilon^i_d+U^i_d=-\varepsilon^i_d$) to $\omega=\varepsilon^f_d$ (and $\varepsilon^f_d+U^f_d=-\varepsilon^f_d$), 
occurs on time scales $|t|\lesssim 1/\Gamma$, occurring
at positive times for the case $t=t_1$
[Fig.~\ref{fig:fig1}(b)], at negative times for the case
$t=t_2$  [Fig.~\ref{fig:fig1}(e)] and at both positive and negative times $-1/\Gamma \lesssim t \lesssim +1/\Gamma$ for the average time spectral function [Figs.~\ref{fig:fig1}(c) and \ref{fig:fig1}(d)]. We note that the shift of the satellite peaks to their final state positions for the average time spectral function occurs in two stages, with half the shift occurring at negative times and the remaining shift occurring at positive times. Another common
feature of all three spectral functions is that, while they all obey the spectral weight sum rule (\ref{eq:sumrule}) exactly, analytically, at all times, and to high accuracy numerically \cite{Note2}, they nevertheless also exhibit regions of negative spectral weight for certain time ranges. This occurs in all cases in the time range where the largest amount of spectral weight is being rearranged, i.e., for $0<t\lesssim +1/\Gamma$ in the case $t=t_1$ [Fig.~\ref{fig:fig1}(b)], at $-1/\Gamma\lesssim t <0$ for the case $t=t_2$ [Fig.~\ref{fig:fig1}(e)], and in the time range $-1/\Gamma \lesssim t \lesssim +1/\Gamma$ for the average time spectral function [Figs.~\ref{fig:fig1}(c) and \ref{fig:fig1}(d)]. These regions of negative spectral weight occur mainly in the frequency range above the satellite peaks in the first case [Fig.~\ref{fig:fig1}(b)], predominantly in the frequency range between the satellite peaks in the last case [Fig.~\ref{fig:fig1}(e)] and both between and above the satellite peaks in the second case [Figs.~\ref{fig:fig1}(c) and \ref{fig:fig1}(d)].

Representative cuts of the spectral function from Fig.~\ref{fig:fig1} at long negative ($tT_{\rm K} =-10^4$) and positive ($tT_{\rm K} =+10^4$)
times as well as at $tT_{\rm K}=0$ are shown in Fig.~\ref{fig:fig2} for all three time references and illustrate the recovery of the initial and final state spectra at long negative/positive times.  At $tT_{\rm K}=0$, one sees that the satellite peaks for the average time
spectral function [Fig.~\ref{fig:fig2}(e)] lies halfway between the initial (vertical dot-dashed lines) and final state (vertical dashed lines) values.

We also note that while the positions of the two satellite peaks acquire their expected final state values by time  $t\gtrsim +1/\Gamma$, or earlier for the case $t=t_2$, their detailed structure continues to vary at longer time scales, reflecting
the drawing of spectral weight from these high-energy satellite peaks to lower energies in the process of  building up the
final state Kondo resonance, which only full develops at the much longer time scale  $t\gtrsim 1/T_{\rm K}$, as we describe next.

\begin{figure}[h]
	\includegraphics[width=0.45\textwidth]{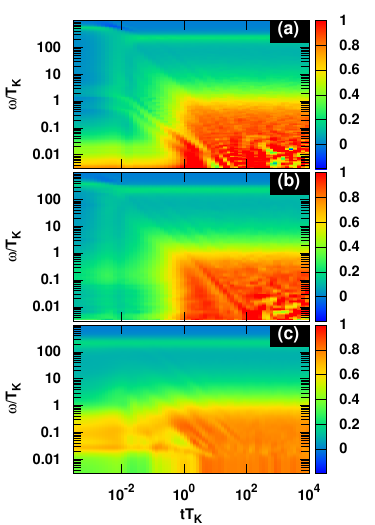}
        \caption{$\pi\Gamma A(\omega>0,t)$ vs $tT_{\rm K}$ at positive times, as in Fig.~\ref{fig:fig1} (right panels), but on a  logarithmic scale for both time and frequency, and for the same quench: (a) $t=t_1$; (b) $t=(t_1+t_2)/2$; (c) $t=t_2$. The 
strong time-dependence of the Kondo resonance around $\omega=0$ is clearly visible in each case, while that of the 
high-energy satellite peak is more clearly resolved on the linear frequency scale of Fig.~\ref{fig:fig1}. 
Signatures of the initial state Kondo resonance of width $T^i_{\rm K}=0.0012T_{\rm K}$ are 
visible in (a) for $t=t_1$, and partially in (b) for $t=(t_1+t_2)/2$ at short times, whereas for $t=t_2$ in (c) the initial state 
Kondo resonance is absent at short times and instead one observes a preformed final state Kondo resonance of width $T_{\rm K}$.
{Results for $t=t_2$ in (c) are from Ref.~\onlinecite{Nghiem2017} and are included here for the purpose of comparison.}
}
    \label{fig:fig3}
\end{figure}
\begin{figure}[h]
  \includegraphics[width=0.45\textwidth]{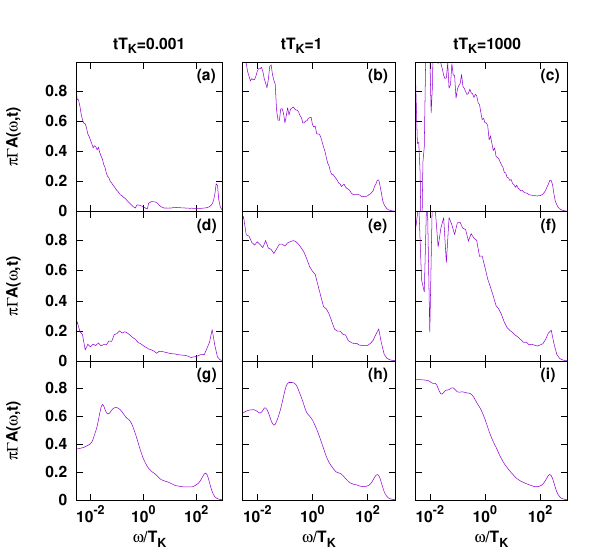}
  \caption 
  {Cuts of the spectral function in Fig.~\ref{fig:fig3} at selected (fixed) times $tT_{\rm K}=0.001$ (left panels), $tT_{\rm K}=1$ (middle panels) and  $tT_{\rm K}=1000$ (right panels). 
    Top panels refer to $t=t_1$, middle panels to $t=(t_1+t_2)/2$ and lower panels to $t=t_2$.
{Results for $t=t_2$ in (g)-(i) are from Ref.~\onlinecite{Nghiem2017} and are included here for the purpose of comparison.}
  }
    \label{fig:fig4}
\end{figure}

The evolution of the Kondo resonance at positive times shows important differences for the spectral functions defined using the three different time references. In order to elucidate these differences, we show in Fig.~\ref{fig:fig3} all three spectral functions at just positive times and on a logarithmic frequency scale in order to better resolve the time evolution of the exponentially narrow Kondo resonance. Representative cuts of the spectral function from Fig.~\ref{fig:fig3} at $tT_{\rm K}=0.001, 1$ and $1000$ are shown in Fig.~\ref{fig:fig4} for all three cases.

We compare first the cases $t=t_1$ [Fig.~\ref{fig:fig3}(a) and Figs.~\ref{fig:fig4}(a)-\ref{fig:fig4}(c)] and
$t=t_2$  [Fig.~\ref{fig:fig3}(c) and Figs.~\ref{fig:fig4}(g)-\ref{fig:fig4}(i)]. Since in the former case [Eq.~(\ref{eq:gf-positive-t1})], time evolution from the initial state only starts at $t=0$, we see in Fig.~\ref{fig:fig3}(a) [and in Fig.~\ref{fig:fig4}(a)] signatures of the initial state Kondo resonance already  at early times $t<10^{-1}/T_{\rm K}$, whose
width is also significantly smaller than that of the final state Kondo resonance.
For $t>10^{-1}/T_{\rm K}$ one sees a crossover to a broader structure which eventually develops into the fully fledged final state Kondo resonance on time scales $t\gtrsim 1/T_{\rm K}$
with a width which is clearly set by the final state Kondo scale $T_{\rm K}$ [see also Figs.~\ref{fig:fig4}(b) and \ref{fig:fig4}(c)]. This evolution is clearly different from that of the spectral function with time taken as $t=t_2$ [ Fig.~\ref{fig:fig3}(c) and lower panels of Fig.~\ref{fig:fig4}]. In the latter, the satellite peaks have already acquired their final state values by $t=0$ and hence,
a structure of width equal to the final state Kondo scale $T_{\rm K}$ is already discernible on this early time scale [Fig.~\ref{fig:fig3}(c) and Fig.~\ref{fig:fig4}(g)]. The subsequent evolution of this structure, or preformed Kondo resonance,
to its fully fledged one, occurs, not via a change in its width as in the case $t=t_1$, but rather by the filling in of the absent spectral weight around the Fermi level at $|\omega|\ll T_{\rm K}$. This occurs on a timescale $t\gtrsim 1/T_{\rm K}$ [Figs.~\ref{fig:fig4}(h) and \ref{fig:fig4}(i)]. For the average time spectral function [Fig.~\ref{fig:fig3}(b) and Figs.~\ref{fig:fig4}(d)-\ref{fig:fig4}(f)], we see that while signatures of the initial state Kondo resonance
are present at early times $t\to 0$, the width of this feature is intermediate between the initial and final state
Kondo scales [Fig.~\ref{fig:fig4}(d)], consistent with the fact that the satellite peaks have only shifted halfway towards their final state values by time $t=0$. The subsequent evolution of the Kondo resonance for average time occurs both via an
increase in its width towards $T_{\rm K}$ (similar to the case $t=t_1$ in Sec.~\ref{sec:gf-times}) and via filling in of states
around the Fermi level in the region $|\omega|\lesssim T_{\rm K}$ on a time scale
$t\gtrsim 1/T_{\rm K}$ [Fig.~\ref{fig:fig4}(e)]. We also notice from Fig.~\ref{fig:fig3}(b), that the transition to the fully developed Kondo resonance occurs rather sharply and within a decade in time on approaching $1/T_{\rm K}$. In contrast,
the Kondo resonance for the cases $t=t_1$ and $t=t_2$ develops over a somewhat wider time range. Finally, in Figs.~\ref{fig:fig3}(a)-\ref{fig:fig3}(c) one clearly sees how the evolution of the satellite peaks to their final state positions, and the associated spectral weight rearrangement, leads to weight being transfered from high to low energies in the process of  building up the final state Kondo resonance [diagonal and vertical stripes, particularly evident in Figs.~\ref{fig:fig3}(a) and \ref{fig:fig3}(b)].

We comment also on the small additional structures within the Kondo resonance which remain
to long times $t=1000/T_{\rm K}$ [Figs.~\ref{fig:fig4}(c), \ref{fig:fig4}(f) and \ref{fig:fig4}(i)].
These have been described elsewhere \cite{Nghiem2017} and are in part due to the use of a Wilson chain in the TDNRG calculations \cite{Rosch2012,Eidelstein2012,Guettge2013} and in part due to the broadening procedure, see Appendix~\ref{sec:appendix-broadening}.

\begin{figure}
  \includegraphics[width=0.45\textwidth]{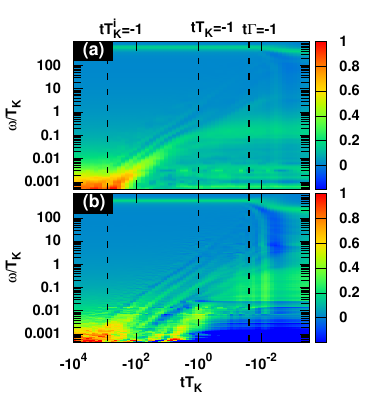}
  \caption{$\pi\Gamma A(\omega>0,t)$ vs $tT_{\rm K}$ at negative times using a logarithmic scale for both time and frequency. 
    (a) Spectral function using $t=(t_1+t_2)/2$, (b) spectral function using $t=t_2$. 
The use of a logarithmic frequency scale allows the decay of the initial state Kondo resonance on the time scale $tT^i_{\rm K}=-1$ and its evolution towards an incipient final state Kondo resonance at $tT_{\rm K}=-1$
          to be clearly seen. 
          The time evolution of the high-energy satellite peaks, also visible, are more clearly seen on the linear frequency scale of Fig.~\ref{fig:fig1}.
          {Results for $t=t_2$ in (b) are from the Supplementary Material of Ref.~\onlinecite{Nghiem2017} and are included here for
            the purpose of comparison.}
        }
    \label{fig:fig5}
\end{figure}

The spectral function at average time exhibits nontrivial time evolution also at negative times. Hence, it is of interest to compare 
this with that of the spectral function with time reference $t=t_2$, which also exhibits nontrivial time evolution at negative times. The comparison is 
shown in Fig.~\ref{fig:fig5} on a logarithmic frequency scale in order to resolve the time evolution of the initial state Kondo resonance. 
We see that in both cases, an initial state Kondo resonance of width $T^i_{\rm K}=0.0012T_{\rm K}$ is present at large negative times $t\to -\infty$. In both 
cases, this initial state Kondo resonance decays on times of order $t=-1/T_{\rm K}^{i}$ and for times $t$ between $-1/T_{\rm K}^{i}$ and $-1/T_{\rm K}$ continues 
to lose spectral weight, with the weight being drawn into a new feature of width $T_{\rm K}$ which can be identified as the incipient final state Kondo resonance whose 
main time evolution occurs at positive times [see Figs.~\ref{fig:fig3}(b) and \ref{fig:fig3}(c)]. We also note that the latter feature, in both cases, draws spectral weight from the decaying initial state 
Kondo resonance, as seen by the diagonal stripes in the figure, and also from the high-energy satellite peaks, as seen by the almost vertical stripes emanating from the high-energy features 
for times $t\gtrsim -1/\Gamma$. 

Summarizing this section, we see that the time evolution of the spectral function clearly depends on whether we chose $t=t_1$, $t=(t_1+t_2)/2$ or $t=t_2$ in its definition.
While the first case only exhibits time evolution for  positive times, the latter cases show nontrivial time evolution also for negative times.
However, all spectral functions exhibit the charge and spin fluctuation time scales $t\approx \pm 1/\Gamma$ and $t\approx 1/T_{\rm K}$
for the evolution of the high- and low-energy features, respectively, and they all exhibit regions of negative spectral density on time scales where the largest spectral weight is being rearranged, while the spectral sum rule is satisfied in each case at all times. In addition, all definitions recover the same equilibrium initial and final state spectral functions in the limits $t\to -\infty$ and $t\to +\infty$, respectively. In the following section, we consider the lesser Green function at the average time $t=(t_1+t_2)/2$, which yields information about the occupied density of states and is closely related to the spectral function at average time and to the time- and energy-resolved photoemission current in pump-probe time-resolved photoemission spectroscopy.

\section{Lesser Green function and time-resolved photoemission currents}
\label{sec:glesser} 
A direct measurement of the time evolution of the single-particle spectral function as a sum over paths of amplitudes in which a particle is added at a certain
time and removed at a later time, is actually not possible experimentally. Instead, one proceeds via time-resolved photoemission spectroscopy  using a pump-probe technique \cite{Perfetti2006,Bovensiepen2012,Eich2017}.
This measures the energy-resolved photoelectron current intensity $I(E,t_{\rm d})$ as a function of the energy of the photoemitted electrons $E$ and the
delay time $t_{\rm d}$ between the probe and the pump pulses. The pump
at time $t=0$ puts the system in a nonequilibrium excited state and corresponds to the quench in our system, while the probe generates a photoelectron current at time $t_d$. {The theory of time-resolved photoemission, relating the intensity $I(E,t_d)$ to Green functions,
involves a number of approximations, see Refs.~ \onlinecite{Freericks2009,Freericks2015,Freericks2017,Randi2017} for details.} For a
Gaussian probe-pulse $s(t)=\exp(-t^2/2\Delta t^2)$ of width $\Delta t$, the photoemission current intensity takes the form \cite{Freericks2009,Freericks2015,Freericks2017,Randi2017}
\begin{align} 
I(E,t_{\rm d})\sim\int d\omega dt\quad N(\omega,t) e^{-\frac{(t-t_d)^2}{\Delta t^2}}e^{-\frac{(\omega-E)^2}{\Delta E^2}}.\label{eq:pes-current}
\end{align} 
Here $N(\omega,t)=\int \frac{d\tau}{2\pi i} e^{i\omega \tau}G^<(t+\frac{\tau}{2},t-\frac{\tau}{2})=G^{<}(\omega,t)/(2\pi i)$ is
the Fourier transform of the lesser Green function defined at the average time, $G^<(t_1,t_2)=i\langle d^{\dagger}_{\sigma}(t_2)d_{\sigma}(t_1)\rangle$ with $t_1=t+{\tau}/{2}$ and $t_2=t-\tau/2$ and $\Delta E = 1/\Delta t$ reflects {the trade-off between the time resolution and the energy resolution, which resembles the quantum mechanical time-energy uncertainty.} Hence, a measurement of $I(E,t_d)$
measures the time-dependent occupied density of states $N(E,t_d)$ convoluted with a Gaussian of width $\Delta t$ in time and a Gaussian of width
$\Delta E=1/\Delta t$ in energy. By analogy to the equilibrium case, where $N(\omega,t)$ reduces to the time-independent
occupied part of the spectral function $f(\omega)A(\omega)$ [see Eq. ~(\ref{eq:fd1})], which can be measured by
standard photoemission spectroscopy, a measurement of $I(E,t_d)$ with time-resolved photoemission 
gives information on the  occupied part of the time-dependent spectral function, see Eq.~(\ref{eq:Aglesser}).

In the following, we first present the result for the lesser Green function at average time within TDNRG (Sec.~\ref{subsec:lgf}), {discussing also its physical structure,} and then use this to
calculate the time-dependent occupied density of states $N(\omega,t)$ in Sec.~\ref{subsec:tr-dos}. In Sec.~\ref{subsec:tr-current} we also
present results for the time-resolved photoemission current $I(E,t_d)$ and investigate the effect 
of using different widths of the Gaussian probe-pulse on the time evolution and observability of spectral features in $N(\omega,t)$.

\subsection{Lesser Green function}
\label{subsec:lgf}
In order to calculate $N(\omega,t)=\frac{G^<(\omega,t)}{2\pi i}$, we require the expression for the lesser Green function at average time within the TDNRG. For positive average time $t$ we find
\begin{align}
&G^<(\omega,t=(t_1+t_2)/2>0)=\nonumber\\
&\sum_{m=m_0}^N\sum_{rsq}^{\notin KK'K''}C^m_{rs}B^m_{sq}\frac{e^{-i(E^m_q-E^m_r)t}-e^{{-2i(\omega+E^m_s-E^m_r)t}}e^{-2\eta t}}{\omega+E^m_s-\frac{E^m_q+E^m_r}{2}-i\eta}\rho^{i\to f}_{qr}(m)\nonumber\\
-&\sum_{m=m_0}^N\sum_{rsq}^{\notin KK'K''}C^m_{rs}B^m_{sq}\frac{e^{-i(E^m_q-E^m_r)t}-e^{{2i(\omega+E^m_s-E^m_q)t}}e^{-2\eta t}}{\omega+E^m_s-\frac{E^m_q+E^m_r}{2}+i\eta}\rho^{i\to f}_{qr}(m)\nonumber\\
+&\sum_{m=m_0}^N\sum_{rsr_1s_1}^{\notin KK'K_1K_1'}C^m_{rs}e^{-2i(\omega-E^m_r+E^m_s)t}e^{-2\eta t}\frac{S^m_{ss_1}\sum_qB^m_{s_1q}\tilde{R}^m_{qr_1}S^m_{r_1r}}{\omega-\frac{E^m_r-E^m_s+E^m_{r_1}-E^m_{s_1}}{2}-i\eta}\nonumber\\
  -&\sum_{m=m_0}^N\sum_{rsr_1s_1}^{\notin KK'K_1K_1'}B^m_{rs}e^{2i(\omega+E^m_r-E^m_s)t}e^{-2\eta t}\frac{S^m_{ss_1}\sum_q\tilde{R}^m_{s_1q}C^m_{qr_1}S^m_{r_1r}}{\omega+\frac{E^m_r-E^m_s+E^m_{r_1}-E^m_{s_1}}{2}+i\eta}
     \label{eq:glesser-positive-t12},
\end{align}
while for negative average time $t$, we find
\begin{align}
&G^<(\omega,t=(t_1+t_2)/2<0)=\nonumber\\
&\sum_{m=m_0}^N\sum_{rs}^{\notin KK'}B^m_{rs}\frac{1-e^{2i(\omega+E^m_r-E^m_s)t}e^{2\eta t}}{\omega+E^m_r-E^m_s-i\eta}\sum_q\tilde{R}^m_{sq}C^m_{qr}\nonumber\\
-&\sum_{m=m_0}^N\sum_{rs}^{\notin KK'}B^m_{rs}\frac{1-e^{-2i(\omega+E^m_r-E^m_s)t}e^{2\eta t}}{\omega+E^m_r-E^m_s+i\eta}\sum_q\tilde{R}^m_{sq}C^m_{qr}\nonumber\\
+&\sum_{m=m_0}^N\sum_{rsr_1s_1}^{\notin KK'K_1K_1'}C^m_{rs}e^{2i(\omega-E^m_{r_1}+E^m_{s_1})t}e^{2\eta t}\frac{S^m_{ss_1}\sum_qB^m_{s_1q}\tilde{R}^m_{qr_1}S^m_{r_1r}}{\omega-\frac{E^m_r-E^m_s+E^m_{r_1}-E^m_{s_1}}{2}-i\eta}\nonumber\\
  -&\sum_{m=m_0}^N\sum_{rsr_1s_1}^{\notin KK'K_1K_1'}B^m_{rs}e^{-2i(\omega+E^m_{r_1}-E^m_{s_1})t}e^{2\eta t}\frac{S^m_{ss_1}\sum_q\tilde{R}^m_{s_1q}C^m_{qr_1}S^m_{r_1r}}{\omega+\frac{E^m_r-E^m_s+E^m_{r_1}-E^m_{s_1}}{2}+i\eta}
     \label{eq:glesser-negative-t12}.
\end{align}
The derivations of these expressions are similar to those for the advanced Green function, which is given in detail in Appendix~\ref{sec:appendix-advanced}.
  
As for the retarded Green functions at average time in Eqs.~(\ref{eq:gf-positive-t12})-(\ref{eq:gf-negative-t12}) of Sec.~\ref{sec:gf-times}, the lesser Green functions here also consist of two types of term: the first two lines of (\ref{eq:glesser-positive-t12}) and (\ref{eq:glesser-negative-t12}) are regular, involving final state (initial state) excitations for $t>0$ ($t<0$), while the last two lines consist of poles at sums of initial and final state excitations (weighted by $e^{-2\eta |t|}$). The latter, decaying as $e^{-2\eta|t|}$ with increasing $t$, describe the decay of initial- and final-state contributions in the limits $t\to+\infty$ and $t\to -\infty$, respectively. In the infinite past, Eq.~(\ref{eq:glesser-negative-t12}) recovers the expression for the lesser Green function of the initial state,
\begin{align}
&G^<(\omega,t\to -\infty)\nonumber\\
=&\sum_{m=m_0}^N\sum_{rs}^{\notin KK'}B^m_{rs}\frac{1}{\omega+E^m_r-E^m_s-i\eta}\sum_q\tilde{R}^m_{sq}C^m_{qr}\nonumber\\
  &-\sum_{m=m_0}^N\sum_{rs}^{\notin KK'}B^m_{rs}\frac{1}{\omega+E^m_r-E^m_s+i\eta}\sum_q\tilde{R}^m_{sq}C^m_{qr}\nonumber\\
=&\sum_{m=m_0}^N\sum_{rs}^{\notin KK'}B^m_{rs}\frac{2i\eta}{(\omega+E^m_r-E^m_s)^2+\eta^2}\sum_q\tilde{R}^m_{sq}C^m_{qr}, \label{eq:glesser-inf-past}
  \end{align}
  while in the infinite future, Eq.~(\ref{eq:glesser-positive-t12}) reduces to the final-state lesser Green function
\begin{align}
&G^<(\omega,t\to +\infty)\nonumber\\
=&\sum_{m=m_0}^N\sum_{rsq}^{\notin KK'K''}C^m_{rs}B^m_{sq}\frac{\delta_{qr}}{\omega+E^m_s-\frac{E^m_q+E^m_r}{2}-i\eta}\rho^{i\to f}_{qr}(m)\nonumber\\
  &-\sum_{m=m_0}^N\sum_{rsq}^{\notin KK'K''}C^m_{rs}B^m_{sq}\frac{\delta_{qr}}{\omega+E^m_s-\frac{E^m_q+E^m_r}{2}+i\eta}\rho^{i\to f}_{qr}(m)\nonumber\\
=&\sum_{m=m_0}^N\sum_{rsq}^{\notin KK'K''}C^m_{rs}B^m_{sq}\frac{2i\eta\delta_{qr}}{(\omega+E^m_s-\frac{E^m_q+E^m_r}{2})^2+\eta^2}\rho^{i\to f}_{qr}(m). \label{eq:glesser-inf-future}
\end{align}
The continuity at $t=0$ is fulfilled as $G^<(\omega,t=0^+)=G^<(\omega,t=0^-)$ can be derived directly from Eqs.(\ref{eq:glesser-positive-t12})-(\ref{eq:glesser-negative-t12}).

\subsection{Results for the time-resolved occupied density of states}
\label{subsec:tr-dos}
\begin{figure}
    \centering 
  	\includegraphics[width=0.482\textwidth]{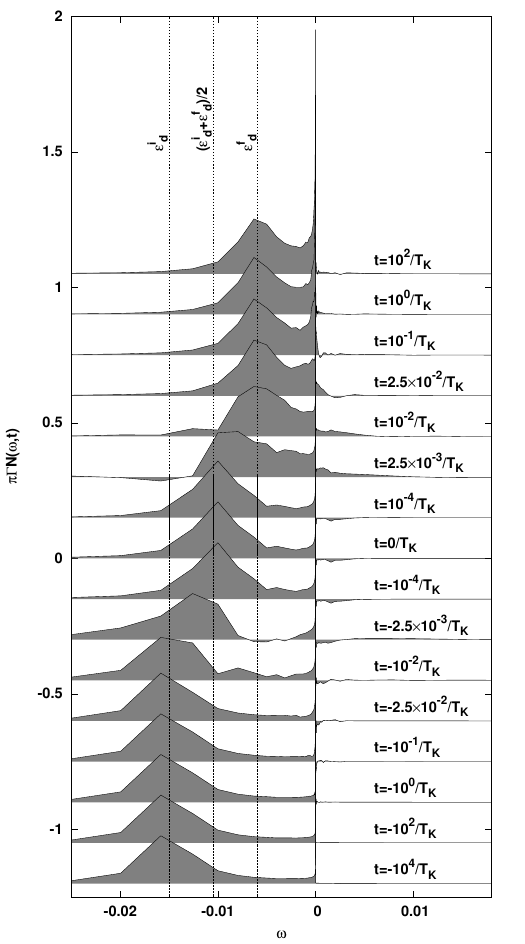}
    \caption{The normalized occupied density of states $\pi\Gamma N(\omega,t)=\Gamma G^{<}(\omega,t)/(2 i)$ vs $\omega$ for different values of the average-time $t$, with the finite-$t$ curves offset vertically by increments of $\pm 0.15$ (for positive/negative times) relative to the $t=0$ curve. The dashed lines mark $\omega=\varepsilon^i_d$, $(\varepsilon^i_d+\varepsilon^f_d)/2$, and $\varepsilon^f_d$. The high-energy satellite-peak starts to shift from $\omega=\varepsilon^i_d$ at $t=-2.5\times 10^{-2}/T_{\rm K}=-1/\Gamma$ to $\omega=(\varepsilon^i_d+\varepsilon^f_d)/2$ at $t=-10^{-4}/T_{\rm K}$, and to shift from $\omega=(\varepsilon^i_d+\varepsilon^f_d)/2$ at $t=10^{-4}/T_{\rm K}$ to $\omega=\varepsilon^f_d$ at $t=2.5\times 10^{-2}/T_{\rm K}=1/\Gamma$. TDNRG parameters as in Fig.~\ref{fig:fig1}.} 
    \label{fig:fig6}
\end{figure}

Figure~\ref{fig:fig6} shows the time evolution of the normalized occupied density of states $\pi\Gamma N(\omega,t) = \Gamma G^<(\omega,t)/2 i$ at selected times from the distant past to the far future for the same quench on the Anderson model as used in the previous sections, i.e., a quench on the symmetric Anderson model from $U_{i}=30\Gamma$ to $U_{f}=12\Gamma$. For an an overview of the behavior
of $\pi\Gamma N(\omega,t)$ at all times, see also Figs.~\ref{fig:fig7}(a)-\ref{fig:fig7}(b) in Sec.~\ref{subsec:tr-current}.

The occupied density of states clearly starts to evolve already at negative times, with the initial state Kondo resonance decaying
on a time scale $tT^i_{\rm K}\gtrsim -1$, i.e., for $tT_{\rm K}\gtrsim -T_{\rm K}/T^{i}_{\rm K}\approx -10^3$ (using $T^i_{\rm K}=0.0012T_{\rm K}$).
This decay is more clearly visible in Fig.~\ref{fig:fig7}(a). In the process, spectral weight from the initial state Kondo resonance and
from the high energy satellite peak is drawn in to form a feature on a scale $T_{\rm K}\gg T^i_{\rm K}$ about the Fermi level,
see for example the curve for $t=-10^{-2}/T_{\rm K}$ or  Fig.~\ref{fig:fig7}(a).
The further time evolution of this feature leads to the build up of the fully developed final state
Kondo resonance at the Fermi level for times $t\gtrsim 1/T_{\rm K}$. In addition to the above low energy changes in
$N(\omega,t)$, which extend to long times of order $1/T_{\rm K}$ or $1/T^i_{\rm K}$,  we also observe large changes
in $N(\omega,t)$ at high energies, occurring mainly on the short time scale $|t| \lesssim 1/\Gamma$:
namely, the high energy satellite peak at $\omega=\varepsilon^i_d$ at $t=-2.5\times 10^{-2}/T_{\rm K}=-1/\Gamma$ shifts first to $\omega=(\varepsilon^i_d+\varepsilon^f_d)/2$ at $t=-10^{-4}/T_{\rm K}=-0.004/\Gamma$, and then from $\omega=(\varepsilon^i_d+\varepsilon^f_d)/2$ at $t=10^{-4}/T_{\rm K}=0.004/\Gamma$ to its final state value $\omega=\varepsilon^f_d$
at $t=2.5\times 10^{-2}/T_{\rm K}=1/\Gamma$ (see vertical dashed lines in Fig.~\ref{fig:fig6}). 

We note that, as for the spectral functions $A(\omega,t)$ defined via the retarded Green function in Sec.~\ref{subsec:spectra-comparisons},
we also observe for $N(\omega,t)$ regions of negative spectral weight in certain time ranges, and particularly in the time range $-1/\Gamma \lesssim t \lesssim +1/\Gamma$ where most of the spectral weight rearrangement takes place as a result of the shift of the satellite peaks from their initial to their final state positions. Such regions of negative spectral weight, for certain time ranges, are also observed in other systems \cite{Jauho1994,Dirks2013,Freericks2009}. Another feature of Fig.~\ref{fig:fig6} is the significant spectral
weight at positive energies and long times ($t\to \infty$), even though the calculation is at $T=0$. We considered two possibilities for this behavior. First, the use of a too large broadening $\eta$ in the Lorentzian broadening procedure for (\ref{eq:glesser-positive-t12}) could
result in a finite spectral weight at $\omega>0$, even at long times. This is because the long tails of Lorentzians can result
in negative energy excitations contributing to the spectrum at $\omega>0$. However, as we show in Appendix~\ref{sec:appendix-broadening},
this is not the case here, as the broadening used gives converged results for $N(\omega,t)$. The extent of the spectral weight at $\omega>0$ for $t\to +\infty$, given that $T=0$, rather suggests that the system has not perfectly equilibrated at long times,
i.e., $N(\omega,t\to +\infty)/A(\omega, t\to\infty)\neq f(\omega)$ where $f(\omega)$ is the equilibrium Fermi function at $T=0$. In Appendix~\ref{sec:appendix-thermalization} we show that instead, $N(\omega,t\to +\infty)/A(\omega, t\to\infty)$ follows, at low frequencies $\omega$, approximately a Fermi function $f_{\rm eff}(\omega)$ with a small effective temperature $T_{\rm eff}\approx T_{\rm K}\Gamma/D $, which is independent of the initial state.  Such an imperfect thermalization at long positive times is expected within the single-quench TDNRG approach \cite{Nghiem2017}. A more precise description of the thermalization at infinite time can be achieved within the multiple-quench TDNRG approach \cite{Nghiem2018}.

\subsection{Results for the time-resolved photoemission current}
\label{subsec:tr-current}

\begin{figure}[t]
    \centering
            \includegraphics[width=0.482\textwidth]{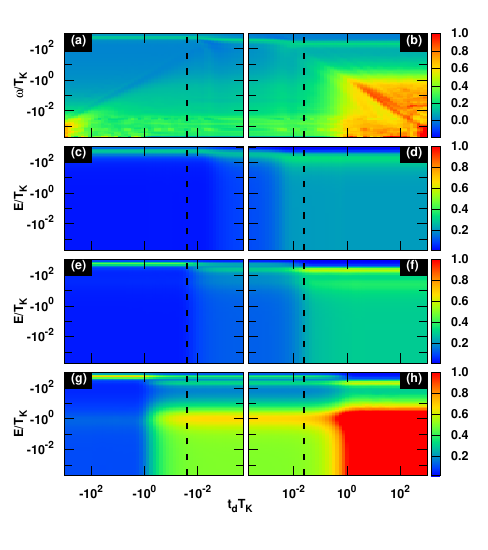}
            \caption{
	Top panels (a) and (b) are $\pi\Gamma N(\omega,t)$ vs time $tT_{\text K}$ 
	and frequency $\omega/T_{\text K}$ on logarithmic scales.
	Lower panels, (c)-(h), are $\pi\Gamma I(E,t_d)$ vs the delay time $t_d T_{\text K}$
	and the energy $E/T_{\text K}$ also on logarithmic scales. {Anderson model quench as in Fig. \ref{fig:fig1}.}
	Panels (c) and (d) are the photoemission current for the short width of probe pulse
	$\Delta t=1/|\varepsilon_d^f| \approx 1/240T_{\text K}$.
	Panels (e) and (f) are the photoemission current for the medium width of probe pulse
	$\Delta t=1/\Gamma \approx 1/40T_{\text K}$.
	Panels (g) and (h) are the photoemission current for the long width of probe pulse
	$\Delta t=1/T_{\text K}$. {Vertical dashed lines indicate $t\Gamma=\pm 1$ in (a) and (b) and $t_d\Gamma=\pm 1$ in the other panels.}}
    \label{fig:fig7}
\end{figure}

\begin{figure}[t]
    \centering
        \includegraphics[width=0.482\textwidth]{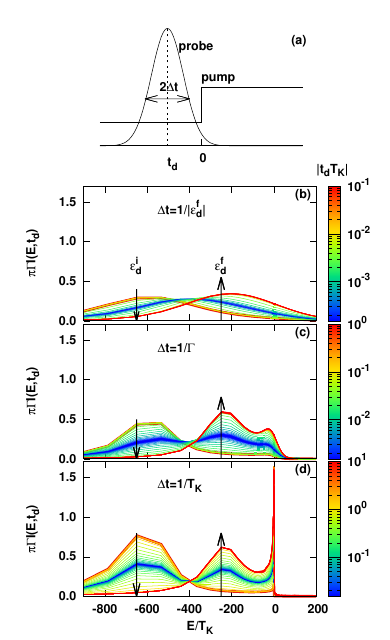}
    \caption{{(a) Sketch of the pump and probe pulses, with $t_d$ the delay time.
	(b)-(d) Cuts of the normalized photoemission currents $\pi\Gamma I(E,t_d)$
        in Fig. \ref{fig:fig7} for specific delay times $t_dT_{\rm K}$, and vs $E/T_{\text K}$ on a linear scale.
	The values of $|t_d T_{\rm K}|$ are given by the color/gray-scale box on the right.
	From (b) to (d), the photoemission currents are calculated with increasing
	width of the probe pulse: (b) $\Delta t=1/|\varepsilon_d^f| \approx 1/240T_{\text K}$,
	(c) $\Delta t=1/\Gamma \approx 1/40T_{\text K}$, and (d) $\Delta t=1/T_{\text K}$. 
	The arrows represent the time evolution of the two peaks at $\varepsilon_d^i$ 
	and $\varepsilon_d^f$ from negative to positive delay times $|t_d T_{\rm K}|$}.}
    \label{fig:fig8}
\end{figure}

The lower panels of Fig.~\ref{fig:fig7} shows the time evolution of the photoemission current intensities $I(E,t_d)$ calculated with three 
different widths of the probe pulse. For comparison, we also show the time evolution of the occupied density of states $N(\omega,t)$ [top panels 
Figs. \ref{fig:fig7}(a) and \ref{fig:fig7}(b)]. We focus here on $I(E,t_d)$ (lower panels) and refer the reader to the description of
Fig. \ref{fig:fig6} given in Sec.~\ref{subsec:tr-dos} for a more detailed description of the time-evolution of $N(\omega,t)$. {We just note,
concerning the latter, that $N(\omega,t)$ in Figs. \ref{fig:fig7}(a) and \ref{fig:fig7}(b) also exhibit signatures of the time scales
$t\Gamma=1$, $tT_{\text K}=1$, and $tT^i_{\text K}=1$, just as in the case of the retarded spectral function $A(\omega,t)$
in Figs. \ref{fig:fig3}(b) and \ref{fig:fig5}(a): namely signatures of the initial state Kondo temperature at $tT_{\rm K}^i\approx -1$ in Fig. \ref{fig:fig7}(a)
and signatures of $\Gamma$ and $T_{\rm K}$ at $t\Gamma\approx 1$ (vertical dashed line) and $tT_{\rm K}=1$, respectively, in Fig.~\ref{fig:fig7}(b).}

Figures \ref{fig:fig7}(c) and \ref{fig:fig7}(d) show the photoemission current intensity $I(E,t_d)$ calculated with 
an ultrashort probe pulse of width $\Delta t=1/|\varepsilon_d^f| \approx 1/240T_{\text K}$ ($3.18\,fs$ for $T_{\rm K}=10\,K$).
The probe-pulse width here is short enough to capture the high energy satellite peak evolving continuously
from $\varepsilon^i_d$ to $\varepsilon^f_d$, but as a result of the low energy-resolution $\Delta E=1/\Delta t \gg T_{\rm K}$
entering the Gaussian in Eq.~(\ref{eq:pes-current}) the low energy Kondo resonance feature in $N(\omega,t)$
can not be resolved. On the other hand, measurements using longer pulse-widths in Figs. \ref{fig:fig7}(e)-\ref{fig:fig7}(h), 
are able to see signatures of the low energy Kondo resonance, but the lack of time resolution does not allow to capture
the detailed time evolution of the high-energy satellite peak from initial to final state positions. Instead, one sees
the initial state peak at long negative delay times and the final state peak at long positive delay times, while at
short delay times signatures of both peaks appear in the photoemission current. This is seen for
the probe-pulse with the longest width $\Delta t=1/T_{\text K}$ ($763.8\,fs$ for $T_{\rm K}=10\,K$), in panels (g) and (h),
where both initial and final state satellite-peaks are present in the signal for delay times ranging
from $t_dT_{\text K}\approx -1$ to $1$, and the low energy Kondo resonance is clearly resolved.

For further insights on the effect of the pulse width on the time evolution of the spectral features
  we examine cuts of $I(E,t_d)$ at specific delay times $t_d$ vs $E/T_{\text K}$ on a linear energy scale in Figs. \ref{fig:fig8}(b)-\ref{fig:fig8}(d).
The pump (quench) and probe pulses are shown schematically in Fig. \ref{fig:fig8}(a).
The photoemission current in Figs. \ref{fig:fig8}(b)-\ref{fig:fig8}(d) is calculated for increasing width of the probe pulses as follows:
(b) $\Delta t=1/|\varepsilon_d^f| \approx 1/240T_{\text K}$ ($3.18\,fs$ for $T_{\rm K}=10\,K$),
(c) $\Delta t=1/\Gamma \approx 1/40T_{\text K}$ ($19.1\,fs$ for $T_{\rm K}=10\,K$), and
(d) $\Delta t=1/T_{\text K}$ ($763.8\,fs$ for $T_{\rm K}=10\,K$).

{
 One sees that for finite delay times around zero, the pump and probe pulses can overlap each other.
Therefore, there is a finite range of delay times, depending on the probe pulse-width, 
such that features of both the initial and final states appear at the same time, as can be observed, for example,
in Figs.~\ref{fig:fig8}(c)-\ref{fig:fig8}(d).}
On the other hand, one sees that the width of the probe pulse acts qualitatively like an effective temperature,
with the smaller the pulse-width, the larger the effective temperature and vice versa. This
again reflects the time-energy uncertainty relation $\Delta E = 1/\Delta t$, since shorter
 pulses have the effect of smearing spectral features $N(\omega,t)$ in the process of extracting $I(E,t_d)$, see
Eq.~(\ref{eq:pes-current}). Therefore, in Fig. \ref{fig:fig8}~(b), the high-energy satellite peaks are low and overbroadened, 
while in Fig. \ref{fig:fig8}~(c) and \ref{fig:fig8}(d), the high-energy satellite peaks are sharper. While the
energy resolution $\Delta E = \Gamma$ in Fig. \ref{fig:fig8}~(c) is not sufficient to fully resolve the Kondo resonance, 
nevertheless, a signal of the Kondo resonance below the Fermi level is clearly seen. 
Finally,  in Fig. \ref{fig:fig8}~(d), when the pulse width is on the scale of the Kondo temperature, 
the Kondo resonance is well resolved.

\section{Conclusions}
\label{sec:conclusions}
In this paper, we investigated several possible definitions for the time-dependent spectral function 
$A(\omega,t)$ of the Anderson impurity model, subject to a sudden quench, and within the TDNRG approach. 
In terms of the retarded (or any other) two-time Green function, $G^r(t_1,t_2)$, one has a choice 
in defining the time $t$ in terms of $t_1$ and/or $t_2$ before carrying out the Fourier transform w.r.t. 
the relative time $t'=t_1-t_2$ to obtain $G^{r}(\omega,t)$ and hence $A(\omega,t)=-{\rm Im}[G^{r}(\omega,t)/\pi]$. 
Choosing $t=t_1$ yields a spectral function which is time-independent for times $t$ before the quench at $t=0$, 
being then identical to the equilibrium initial state spectral function, while having a nontrivial time evolution 
at positive times after the quench. This spectral function appears in the context of time-dependent transport 
through quantum dots with time-dependent parameters \cite{Jauho1994}, but is not a directly measurable observable 
in that context, since it only appears in expressions for transient currents. 
The choice $t=t_2$ \cite{Anders2008b,Nghiem2017}, motivated by applications for extracting steady state nonequilibrium 
spectral functions \cite{Anders2008a}, exhibits nontrivial time evolution at both negative and positive times \cite{Nghiem2017}.  
The choice $t=\frac{t_1+t_2}{2}$ results in a time-dependent spectral function $A(\omega,t)$ which is close to that measured in
time-resolved photoemisssion spectroscopy, which measures the time-dependent occupied density of states
$N(\omega,t) = G^<(\omega,t)/(2\pi i)$, which makes up a part of the average time spectral function [see Eq.~(\ref{eq:Aglesser})]. 
In the context of the experiment, the average time here is identified as the delay time between the pump and the probe pulses.
For the quench that we studied in detail, in which the Coulomb interaction in the symmetric Anderson model is reduced 
from $U_i = 30\Gamma$ in the initial state to $U_f=12\Gamma$ in the final state, we find that, 
in all cases, the final state Kondo resonance in $A(\omega,t)$ is only fully developed for times $t\gtrsim 1/T_{\rm K}$, 
while the largest rearrangement of spectral weight, associated with the high-energy satellite peaks shifting 
from their initial to final state values, occurs on a time $|t|$ of order $1/\Gamma$. However, whereas this 
shift occurs largely around $t=-1/\Gamma$ for the choice $t=t_2$, and largely around $t=+1/\Gamma$ for the choice
$t=t_1$, for the average time spectral function, it occurs in two stages between $t=-1/\Gamma$ and $t=0$ and
between $t=0$ and $t=+1/\Gamma$. 

In addition to deriving expressions for $A(\omega,t)=-{\rm Im}[G^{r}(\omega,t)/\pi]$ for different time references, we
also derived expressions within TDNRG for the advanced , lesser and greater Green functions for the same time references.
This allowed us to explicitly verify that for average times $[G^a(\omega,t)]^*=G^r(\omega,t)$ and that $G^>(\omega,t)$ and $G^<(\omega,t)$ are
purely imaginary, properties that allow a real time-dependent spectral function to be defined as in equilibrium via $A(\omega,t)=\frac{i}{2\pi}\Big[G^r(\omega,t)-G^a(\omega,t)\Big]$ as well as via Eqs.~(\ref{eq:Agagr})-(\ref{eq:Aglesser}).
In contrast, the above properties are not generally satisfied for the other choices of time reference, for which the
definition in terms of the imaginary part of the retarded Green functions is more appropriate. Ultimately, however,
the experimental context dictates which definition applies.

We investigated the average time lesser Green function, which yields the time-dependent occupied density of states 
$N(\omega,t) = G^<(\omega,t)/(2\pi i)$, which in equilibrium reduces to $f(\omega)A(\omega)$, and which
is closely related to the photoemission current $I(E,t_d)$ measured in time-resolved photoemisssion spectroscopy 
[Eq.~(\ref{eq:pes-current})]. $N(\omega,t)$ was also found to have a nontrivial time evolution at both positive and
negative average times as for the spectral function with $t=(t_1+t_2)/2$. While the main spectral weight in $N(\omega,t)$ at $T=0$, was found to be
below the Fermi energy at all times, a small occupation of states above the Fermi level, which persisted to infinite times,
was also found.  
We found that at low frequencies $\omega$ close to the Fermi level an effective Fermi function
$f_{\rm eff}(\omega)=N(\omega,t\to +\infty)/A(\omega, t\to\infty)$ with a small
effective temperature $T_{\rm eff}\approx \Gamma T_{\rm K}/D$, 
independent of the initial state, was consistent with the
data. This imperfect thermalization within the single quench TDNRG approach can be attributed to the discrete
Wilson chain representation of the conduction electron bath \cite{Rosch2012} and can be reduced within a multiple-quench
TDNRG approach \cite{Nghiem2018}.

Finally, in terms of the application of our results to time-resolved photoemission spectroscopy, we
calculated the photoemission current $I(E,t_d)$ from the occupied density of states $N(\omega,t)$ via Eq.~(\ref{eq:pes-current}),
and investigated the observability and the time evolution of spectral features in the photoemission current for Gaussian probe
pulses of different widths. While ultrashort probe pulses yield better time-resolution for the high energy features at early
times, they also yield less energy-resolution and can miss features close to the Fermi energy. Calculations with
three different values of pulse widths inversely proportional to the three relevant energy scales
 $\epsilon^f_d$, $\Gamma$, and $T_{\text K}$ exhibit different behaviour of the photoemission current.
For the measurements with an ultrashort pulse $\Delta t=1/|\epsilon^f_d|$, having, therefore, high time-resolution, 
the photoemission current can capture as a function of the delay time the fast evolution of the high-energy satellite
peak for times close to the time of the quench ($t=0$).
For a pulse with intermediate width $\Delta t=1/ \Gamma$, the photoemission current does not capture the
fast evolution of the high energy satellite peak in detail, but the energy resolution is high enough to start seeing
a signal of the Kondo resonance around the Fermi level $E=0$.
For long probe pulses  $\Delta t=1/T_{\text K}$, therefore having high energy-resolution, the continuous evolution of the high energy
satellite peaks from initial to final state values at short times, cannot be resolved, but the low energy Kondo resonance is
clearly resolved. The above results and insights could be useful for future studies of the time evolution of the Kondo resonance
with time-resolved photoemission spectroscopy. 

Since the TDNRG expressions for the nonequilibrium Green functions presented in this paper hold for general local operators $\hat{B}$ and $\hat{C}$, they
can easily be  generalized to other time-dependent dynamical quantities, e.g.,  to time-dependent dynamical susceptibilities. {The latter can then be used in applications to time-resolved optical conductivity spectroscopy.}

\begin{acknowledgments}
H. T. M. Nghiem acknowledges the support by Vietnam National Foundation for Science and Technology Development (NAFOSTED) under grant number 103.2-2017.353. {Useful discussions with J. K. Freericks are acknowledged.}
We acknowledge support by the Deutsche Forschungsgemeinschaft via the ``Research Training Group 1995'' and supercomputer support by the John von Neumann institute for Computing (J\"ulich).
\end{acknowledgments}

\appendix
\section{Advanced Green function}
\label{sec:appendix-advanced}
The advanced Green function is defined as
\begin{align}
G^a(t_1,t_2)=i\theta(t_2-t_1)\langle\{\hat{B}(t_1),\hat{C}(t_2)\}\rangle
\end{align}
and is transformed into
\begin{align}
G^a(t,\tau)=i\theta(-\tau)\langle\{\hat{B}(t+\tau/2),\hat{C}(t-\tau/2)\}\rangle
\end{align}
with $t=(t_1+t_2)/2$ and $\tau=t_1-t_2$ the average and relative times, respectively.

\begin{widetext}
\subsubsection{Positive time $t>0$}
We have
\begin{equation}
G^a(t,\tau)=\begin{dcases}
     i\operatorname{Tr}\{\hat{\rho}[e^{iH_f(t+\tau/2)}B e^{-iH_f(t+\tau/2)},e^{iH_f(t-\tau/2)}\hat{C}e^{-iH_f(t-\tau/2)}]_+\} & \text{if } -2t<\tau<0;\\
i\operatorname{Tr}\{\hat{\rho}[e^{iH_i(t+\tau/2)}B e^{-iH_i(t+\tau/2)},e^{iH_f(t-\tau/2)}\hat{C}e^{-iH_f(t-\tau/2)}]_+\} &  \text{if } \tau\leq -2t;\\
0 &  \text{otherwise}.
   \end{dcases}
\end{equation}
Denoting the first and second lines of the above expression by $G^-_{BC}(t,\tau)$ and $G^+_{BC}(t,\tau)$, respectively, we have for $G^-_{BC}(t,\tau)$
\begin{align}
G^-(t,\tau)=&i\operatorname{Tr}\{e^{-iH_f(t-\tau/2)}\hat{\rho}e^{iH_f(t-\tau/2)}[e^{iH_f\tau}B e^{-iH_f\tau},\hat{C}]_+\} \nonumber\\
=&i\sum_{l_1e_1m_1}\sum_{l_2e_2m_2}\sum_{l_3e_3m_3} {_f}\langle l_1e_1m_1|e^{-iH_f(t-\tau/2)}\hat{\rho}e^{iH_f(t-\tau/2)}|l_2e_2m_2\rangle_f\nonumber\\
&\times ({_f}\langle l_2e_2m_2|e^{iH_f\tau}B e^{-iH_f\tau}|l_3e_3m_3\rangle_f {_f}\langle l_3e_3m_3|\hat{C} |l_1e_1m_1\rangle_f\nonumber\\
&+{_f}\langle l_2e_2m_2|\hat{C}|l_3e_3m_3\rangle_f {_f}\langle l_3e_3m_3| e^{iH_f\tau}B e^{-iH_f\tau}|l_1e_1m_1\rangle_f)\nonumber\\
=&i\sum_{me}\sum_{rsq}^{\notin KK'K''} {_f}\langle rem|e^{-iH_f(t-\tau/2)}\hat{\rho}e^{iH_f(t-\tau/2)}|sem\rangle_f\nonumber\\
&\times ({_f}\langle sem|e^{iH_f\tau}B e^{-iH_f\tau}|qem\rangle_f {_f}\langle qem|\hat{C} |rem\rangle_f+{_f}\langle sem|\hat{C}|qem\rangle_f {_f}\langle qem| e^{iH_f\tau}B e^{-iH_f\tau}|rem\rangle_f)\nonumber\\
=&i\sum_{m}\sum_{rsq}^{\notin KK'K''} {\rho}^{i\to f}_{rs}(m)e^{i(E^m_s-E^m_r)(t-\tau/2)} (B^m_{sq} e^{i(E^m_s-E^m_q)\tau} C^m_{qr}+C^m_{sq}  B^m_{qr}e^{i(E^m_q-E^m_r)\tau})\nonumber\\
=&i\sum_{m}\sum_{rsq}^{\notin KK'K''} {\rho}^{i\to f}_{rs}(m)e^{i(E^m_s-E^m_r)t} (B^m_{sq} e^{i[(E^m_s+E^m_r)/2-E^m_q]\tau} C^m_{qr}+C^m_{sq}  B^m_{qr}e^{i[E^m_q-(E^m_r+E^m_s)/2]\tau} ),
\end{align}
{in which we use the identity \cite{Weymann2015}
\begin{align}
\sum_{l_{1}e_{1}m_{1}}\sum_{l_{2}e_{2}m_{2}}\sum_{l_{3}e_{3}m_{3}}=\sum_m\sum_{e_1e_{2}e_{3}}\sum_{rsq}^{\notin KK'K''}\label{eq:identity-multiple-shells}
\end{align}
to obtain the third line in the above equation.}
While for $G^+_{BC}(t,\tau)$ we have
\begin{align}
G^+(t,\tau)=&i\operatorname{Tr}\{[\hat{\rho},e^{iH_i(t+\tau/2)}\hat{B}e^{-iH_i(t+\tau/2)}]_+ e^{iH_f(t-\tau/2)}\hat{C} e^{-iH_f(t-\tau/2)}\} \nonumber\\
=&i\sum_{l_1e_1m_1}\sum_{l_2e_2m_2}\sum_{l_3e_3m_3}\sum_{l_4e_4m_4}{_f}\langle l_1e_1m_1|l_2e_2m_2\rangle_i\nonumber\\
&\times{_i}\langle l_2e_2m_2|[\hat{\rho},e^{iH_i(t+\tau/2)}\hat{B}e^{-iH_i(t+\tau/2)}]_+|l_3e_3m_3\rangle_i {_i}\langle l_3e_3m_3 |l_4e_4m_4\rangle_f {_f}\langle l_4e_4m_4|e^{iH_f(t-\tau/2)}\hat{C} e^{-iH_f(t-\tau/2)}|l_1e_1m_1\rangle_f \nonumber\\
=&i\sum_{me}\sum_{rsr_1s_1}^{\notin KK'K_1K_1'}{_f}\langle rem|r_1em\rangle_i\nonumber\\
&\times{_i}\langle r_1em|[\hat{\rho},e^{iH_i(t+\tau/2)}\hat{B}e^{-iH_i(t+\tau/2)}]_+ |s_1em\rangle_i {_i}\langle s_1em|sem\rangle_f {_f}\langle sem|e^{iH_f(t-\tau/2)}\hat{C} e^{-iH_f(t-\tau/2)}|rem\rangle_f \nonumber\\
=&i\sum_{m}\sum_{rsr_1s_1}^{\notin KK'K_1K_1'}S^m_{rr_1} \sum_q(B^m_{r_1q}\tilde{R}^m_{qs_1}+\tilde{R}^m_{r_1q}B^m_{qs_1}) e^{i(E^m_{r_1}-E^m_{s_1})(t+\tau/2)} S^m_{s_1s}C^m_{sr} e^{i(E^m_s-E^m_r)(t-\tau/2)}\nonumber\\
=&i\sum_{m}\sum_{rsr_1s_1}^{\notin KK'K_1K_1'}S^m_{rr_1} \sum_q(B^m_{r_1q}\tilde{R}^m_{qs_1}+\tilde{R}^m_{r_1q}B^m_{qs_1}) e^{i(E^m_s-E^m_r+E^m_{r_1}-E^m_{s_1})t} S^m_{s_1s}C^m_{sr} e^{i(E^m_r-E^m_s+E^m_{r_1}-E^m_{s_1})\tau/2},
\end{align}
{with $\tilde{R}$ defined as follows
\begin{equation}
\tilde{R}^m_{rs}=\begin{dcases}
     R^m_{kk'} & \text{if } r=k \in K\text{ and } s=k'\in K';\\
     (w_{m+1}\frac{e^{-\beta E_l^{m+1}}}{Z_{m+1}}\Big) & \text{if } r=s=l\in D;\\
     0 & \text{otherwise}
   \end{dcases}
\end{equation}
and
\begin{equation}
R^m_{kk'}=\begin{dcases}
     0 & \text{if } m=N;\\
\sum_{l\alpha_{m+1}}A^{\alpha_{m+1}}_{kl}\Big(w_{m+1}\frac{e^{-\beta E_l^{m+1}}}{Z_{m+1}}\Big)A^{\alpha_{m+1}\dagger}_{lk'}
+\sum_{k_1k'_1\alpha_{m+1}}A^{\alpha_{m+1}}_{kk_1}R_{k_1k'_1}^{m+1}A^{\alpha_{m+1}\dagger}_{k'_1k'} & \text{otherwise}\label{eq:RFDMrecur},
   \end{dcases}
\end{equation}}
{where the weights $w_m$ in (\ref{eq:RFDMrecur}) are the same as those in the expression (\ref{eq:fdm-rho})
  for the full density matrix of the initial state.}
Fourier transform the resulting Green function we obtain
\begin{align}
&G^a(\omega,t)=\int^{0}_{-2t}d\tau e^{i(\omega-i\eta)\tau}G^-(t,\tau)+\int^{-2t}_{-\infty}d\tau e^{i(\omega-i\eta)\tau}G^+(t,\tau)\nonumber\\
=&\sum_{m}\sum_{rsq}^{\notin KK'K''}  {\rho}^{i\to f}_{rs}(m)e^{i(E^m_s-E^m_r)t} \Big[\frac{B^m_{sq} C^m_{qr}}{\omega+(E^m_s+E^m_r)/2-E^m_q-i\eta}(1- e^{-2i[\omega+(E^m_s+E^m_r)/2-E^m_q-i\eta]t} )\nonumber\\
&\hspace{11em}+\frac{C^m_{sq}  B^m_{qr}}{\omega-(E^m_s+E^m_r)/2+E^m_q-i\eta}(1-e^{-2i[\omega-(E^m_s+E^m_r)/2+E^m_q-i\eta]t})\Big]\nonumber\\
&+\sum_{m}\sum_{rsr_1s_1}^{\notin KK'K_1K_1'}\frac{S^m_{s_1s}C^m_{sr}S^m_{rr_1} e^{i(E^m_s-E^m_r+E^m_{r_1}-E^m_{s_1})t} }{\omega+(E^m_r-E^m_s)/2-(E^m_{s_1}-E^m_{r_1})/2-i\eta} \sum_q(B^m_{r_1q}\tilde{R}^m_{qs_1}+\tilde{R}^m_{r_1q}B^m_{qs_1}) e^{-2i(\omega+(E^m_r-E^m_s)/2-(E^m_{s_1}-E^m_{r_1})/2-i\eta)t} ,\nonumber
\end{align}
which can be rewritten as
\begin{align}
G^a(\omega,t)=&\sum_{m}\sum_{rsq}^{\notin KK'K''}  {\rho}^{i\to f}_{rs}(m)\Big[\frac{B^m_{sq} C^m_{qr}}{\omega+(E^m_s+E^m_r)/2-E^m_q-i\eta}(e^{i(E^m_s-E^m_r)t} - e^{-2i[\omega+E^m_r-E^m_q-i\eta]t} )\nonumber\\
&\hspace{7em}+\frac{C^m_{sq}  B^m_{qr}}{\omega-(E^m_s+E^m_r)/2+E^m_q-i\eta}(e^{i(E^m_s-E^m_r)t} -e^{-2i[\omega-E^m_s+E^m_q-i\eta]t})\Big]\nonumber\\
&+\sum_{m}\sum_{rsr_1s_1}^{\notin KK'K_1K_1'}\frac{S^m_{s_1s}C^m_{sr}S^m_{rr_1} e^{-2i(\omega+(E^m_r-E^m_s)-i\eta)t} }{\omega+(E^m_r-E^m_s)/2-(E^m_{s_1}-E^m_{r_1})/2-i\eta} \sum_q(B^m_{r_1q}\tilde{R}^m_{qs_1}+\tilde{R}^m_{r_1q}B^m_{qs_1}).
\end{align}
Since $\hat{B}\equiv d$ and $\hat{C}\equiv d^{\dagger}$ it follows that $B^m_{sq}=C^m_{qs}$. We also have that ${\rho}^{i\to f}_{rs}(m)={\rho}^{i\to f}_{sr}(m)$, and $\tilde{R}^m_{s_1q}=\tilde{R}^m_{qs_1}$, therefore we can rewrite the above expression as
\begin{align}
G^a(\omega,t)=&\sum_{m}\sum_{rsq}^{\notin KK'K''}  {\rho}^{i\to f}_{rs}(m)\Big[\frac{B^m_{sq} C^m_{qr}}{\omega+(E^m_s+E^m_r)/2-E^m_q-i\eta}(e^{-i(E^m_s-E^m_r)t} - e^{-2i[\omega+E^m_s-E^m_q-i\eta]t} )\nonumber\\
&\hspace{7em}+\frac{C^m_{sq}  B^m_{qr}}{\omega-(E^m_s+E^m_r)/2+E^m_q-i\eta}(e^{-i(E^m_s-E^m_r)t} -e^{-2i[\omega-E^m_r+E^m_q-i\eta]t})\Big]\nonumber\\
&+\sum_{m}\sum_{rsr_1s_1}^{\notin KK'K_1K_1'}\frac{S^m_{rr_1}B^m_{rs}S^m_{ss_1} e^{-2i(\omega+(E^m_r-E^m_s)-i\eta)t} }{\omega+(E^m_r-E^m_s)/2-(E^m_{s_1}-E^m_{r_1})/2-i\eta} \sum_q(\tilde{R}^m_{s_1q}C^m_{qr_1}+C^m_{s_1q}\tilde{R}^m_{qr_1}),\label{eq:gfa-positive-t12}
\end{align}
in which we interchanged $r$ and $s$ in the first term. Comparing the expression of $G^a(\omega,t)$ in Eq.~(\ref{eq:gfa-positive-t12}) with that of $G^r(\omega,t)$ in Eq.~(\ref{eq:gf-positive-t12}), we can easily show that $[G^a(\omega,t>0)]^*=G^r(\omega,t>0)$.
\subsubsection{Negative time $t<0$}
We have
\begin{equation}
G^a_{BC}(t,\tau)=\begin{dcases}
     i\operatorname{Tr}\{\hat{\rho}[e^{iH_i(t+\tau/2)}\hat{B} e^{-iH_i(t+\tau/2)},e^{iH_i(t-\tau/2)}\hat{C}e^{-iH_i(t-\tau/2)}]_+\} &\text{if } 2t<\tau\leq 0;\\
i\operatorname{Tr}\{\hat{\rho}[e^{iH_i(t+\tau/2)}\hat{B} e^{-iH_i(t+\tau/2)},e^{iH_f(t-\tau/2)}\hat{C}e^{-iH_f(t-\tau/2)}]_+\} & \text{if } \tau<2t  ;\\
0 &  \text{otherwise}.
   \end{dcases}
\end{equation}
Denoting the first and second lines of the above expression by $G^-_{BC}(t,\tau)$ and $G^+_{BC}(t,\tau)$, respectively, we have for $G^-_{BC}(t,\tau)$
 \begin{align}
G^-(t,\tau)=& i\operatorname{Tr}\{\hat{\rho}[e^{iH_i\tau}\hat{B} e^{-iH_i\tau},\hat{C}]_+\}\nonumber\\
=& i\operatorname{Tr}\{e^{iH_i\tau}B e^{-iH_i\tau}[\hat{C},\hat{\rho}]_+\}\nonumber\\
=& i\sum_{l_1e_1m_1}\sum_{l_2e_2m_2}{_i}\langle l_1e_1m_1|e^{iH_i\tau}\hat{B} e^{-iH_i\tau}|l_2e_2m_2\rangle_i {_i}\langle l_2e_2m_2|[\hat{C},\hat{\rho}]_+|l_1e_1m_1\rangle_i\nonumber\\
=& i\sum_{me}\sum_{rs}^{\notin KK''} {_i}\langle rem|e^{iH_i\tau}\hat{B} e^{-iH_i\tau}|sem\rangle_i {_i}\langle rem|[\hat{C},\hat{\rho}]_+|sem\rangle_i\nonumber\\
=& i\sum_{m}\sum_{rs}^{\notin KK'} B^m_{rs} e^{i(E^m_r-E^m_s)\tau}\sum_q (C^m_{sq}\tilde{R}^m_{qr}+\tilde{R}^m_{sq}C^m_{qr}),
\end{align}
while the expression for $G^+_{BC}(t,\tau)$ is similar to that for the case of $t>0$ 
\begin{align}
G^+(t,\tau)
=&i\sum_{m}\sum_{rsr_1s_1}^{\notin KK'K_1K_1'}S^m_{rr_1} \sum_q(B^m_{r_1q}\tilde{R}^m_{qs_1}+\tilde{R}^m_{r_1q}B^m_{qs_1}) e^{i(E^m_{r_1}-E^m_{s_1})(t+\tau/2)} S^m_{s_1s}  C^m_{sr} e^{i(E^m_s-E^m_r)(t-\tau/2)}\\
=&i\sum_{m}\sum_{rsr_1s_1}^{\notin KK'K_1K_1'}S^m_{rr_1} \sum_q(B^m_{r_1q}\tilde{R}^m_{qs_1}+\tilde{R}^m_{r_1q}B^m_{qs_1}) e^{i(E^m_s-E^m_r+E^m_{r_1}-E^m_{s_1})t} S^m_{s_1s}  C^m_{sr} e^{i(E^m_r-E^m_s+E^m_{r_1}-E^m_{s_1})\tau/2}.
\end{align}
Fourier-transforming the resulting Green function gives
\begin{align}
&G^a(\omega,t)=\int_{2t}^{0}d\tau e^{i(\omega-i\eta)\tau}G^-(t,\tau)+\int ^{2t}_{-\infty}d\tau e^{i(\omega-i\eta)\tau}G^+(t,\tau)\nonumber\\
=&\sum_{m}\sum_{rs}^{\notin KK'}\frac{B^m_{rs}}{\omega+E^m_r-E^m_s-i\eta} (1-e^{2i(\omega+E^m_r-E^m_s+i\eta)t})\sum_q (C^m_{sq}\tilde{R}^m_{qr}+\tilde{R}^m_{sq}C^m_{qr})\nonumber\\
&+\sum_{m}\sum_{rsr_1s_1}^{\notin KK'K_1K_1'}\frac{S^m_{s_1s}C^m_{sr}S^m_{rr_1} e^{i(E^m_s-E^m_r+E^m_{r_1}-E^m_{s_1})t} }{\omega+(E^m_r-E^m_s)/2-(E^m_{s_1}-E^m_{r_1})/2-i\eta} \sum_q(B^m_{r_1q}\tilde{R}^m_{qs_1}+\tilde{R}^m_{r_1q}B^m_{qs_1}) e^{2i(\omega+(E^m_r-E^m_s)/2-(E^m_{s_1}-E^m_{r_1})/2-i\eta)t} \nonumber\\
=&\sum_{m}\sum_{rs}^{\notin KK'}\frac{B^m_{rs}}{\omega+E^m_r-E^m_s-i\eta} (1-e^{2i(\omega+E^m_r-E^m_s-i\eta)t})\sum_q (C^m_{sq}\tilde{R}^m_{qr}+\tilde{R}^m_{sq}C^m_{qr})\nonumber\\
&+\sum_{m}\sum_{rsr_1s_1}^{\notin KK'K_1K_1'}\frac{S^m_{s_1s}C^m_{sr}S^m_{rr_1}  e^{2i(\omega-(E^m_{s_1}-E^m_{r_1})-i\eta)t} }{\omega+(E^m_{r}-E^m_{s})/2-(E^m_{s_1}-E^m_{r_1})/2-i\eta} \sum_q(B^m_{r_1q}\tilde{R}^m_{qs_1}+\tilde{R}^m_{r_1q}B^m_{qs_1})\nonumber\\
=&\sum_{m}\sum_{rs}^{\notin KK'}\frac{B^m_{rs}}{\omega+E^m_r-E^m_s-i\eta} (1-e^{2i(\omega+E^m_r-E^m_s-i\eta)t})\sum_q (C^m_{sq}\tilde{R}^m_{qr}+\tilde{R}^m_{sq}C^m_{qr})\nonumber\\
&+\sum_{m}\sum_{rsr_1s_1}^{\notin KK'K_1K_1'}\frac{S^m_{rr_1}B^m_{r_1s_1}S^m_{s_1s}  e^{2i(\omega-(E^m_{s}-E^m_{r})-i\eta)t} }{\omega+(E^m_{r_1}-E^m_{s_1})/2-(E^m_{s}-E^m_{r})/2-i\eta} \sum_q(\tilde{R}^m_{sq}C^m_{qr}+C^m_{sq}\tilde{R}^m_{qr}).\label{eq:gfa-negative-t12}
\end{align}
Comparing the above expression with $G^r(\omega,t)$ in Eq.~\ref{eq:gf-negative-t12}, one can see that $[G^a(\omega,t<0)]^*=G^r(\omega,t<0)$. In addition, comparing Eq.~(\ref{eq:gfa-positive-t12}) with Eq.~(\ref{eq:gfa-negative-t12}), we see that the continuity condition $G^a(\omega,t\to 0^+)=G^a(\omega,t\to 0^-)$ is also satisfied.

\section{Advanced, lesser and greater Green functions}
\label{sec:appendix-gf-list}
We  list here the TDNRG expressions for the advanced,
lesser and greater Green functions for all reference times,
complementing those for the retarded Green function and lesser Green
function at average time, which have been given in the main text. The
derivations of these expressions are similar those given for the average
time advanced Green function in Appendix~\ref{sec:appendix-advanced} and the
retarded Green function for $t=t_2$ \cite{Note2}.
\subsection{Advanced Green function}
In the case that $t=t_1$,
\begin{align}
&G^a(\omega,t>0)
=\sum_{m=m_0}^N\sum_{rsq}^{\notin KK'K''}\rho_{sr}^{i\to f}(m)e^{i(E_{r}^{m}-E_{s}^{m})t}\Big(\frac{B^m_{rq}C^m_{qs}}{\omega+E^m_s-E^m_q-i\eta}
+\frac{C^m_{rq} B^m_{qs}}{\omega+E^m_q-E^m_r-i\eta}\Big),\label{eq:gfa-positive-times}
\end{align}

\begin{align}
G^a(\omega,t<0)=&\sum_m\Big[\sum_{rs}^{\notin KK'} \frac{B^m_{rs}}{\omega+E^m_r-E^m_s-i\eta}(1-e^{i(\omega+E^m_r-E^m_s-i\eta)t})\sum_{q}(C^m_{sq}\tilde{R}^m_{qr}+\tilde{R}^m_{sq}C^m_{qr})\nonumber\\
&+\sum_{rsr_1s_1}^{\notin KK'K_1K'_1} S^m_{rr_1}\frac{C^m_{r_1s_1}}{\omega+E^m_{s_1}-E^m_{r_1}-i\eta} S^m_{s_1s}e^{i(\omega+E^m_s-E^m_r-i\eta)t}\sum_{q}(B^m_{sq}\tilde{R}^m_{qr}+\tilde{R}^m_{sq}B^m_{qr})\Big]
.\label{eq:gfa-negative-times}
\end{align}

In the case that $t=t_2$,
\begin{align}
G^a(\omega,t>0)
=&\sum_{m=m_0}^N\Big\{\sum_{rsq}^{\notin KK'K''}\Big[C^m_{rs}\rho_{sq}^{i\to f}(m)e^{-i(E_{q}^{m}-E_{s}^{m})t}+ \rho_{rs}^{i\to f}(m)e^{-i(E_{s}^{m}-E_{r}^{m})t}C^m_{sq}\Big]\frac{B^m_{qr}}{\omega+E^m_q-E^m_r-i\eta}(1-e^{-i(\omega+E^m_q-E^m_r-i\eta)t})\nonumber\\
+&\sum_{rsr_1s_1}^{\notin KK'K_1K'_1} S^m_{rr_1}C^m_{r_1s_1}e^{-i(\omega-E^m_{r_1}+E^m_{s_1}-i\eta)t}S^m_{s_1s}\frac{\sum_{q}(B^m_{sq}\tilde{R}^m_{qr}+\tilde{R}^m_{sq}B^m_{qr})}{\omega-E^m_{r}+E^m_{s}-i\eta}\Big\}\label{eq:gfa-positive-t2},
\end{align}
while $G^a(\omega,t<0)$ is time independent, and exactly equals the
advanced Green function of the initial state for the same reason that
$G^r(\omega,t=t_1<0)$ is time-independent [see discussion preceding Eq.~(\ref{eq:gf-negative-t1})]. 
\subsection{Lesser Green function}
In the case that $t=t_1$
\begin{align}
G^<(\omega,t>0)
=&\sum_{m=m_0}^N\sum_{rsq}^{\notin KK'K''}\Big(\frac{C^m_{rs}B^m_{sq}}{\omega-E^m_r+E^m_s-i\eta}
-\frac{C^m_{rs} B^m_{sq}}{\omega-E^m_r+E^m_s+i\eta}\Big)
\rho_{qr}^{i\to f}(m)e^{i(E_{r}^{m}-E_{q}^{m})t}\nonumber\\
&+\sum_{m=m_0}^N\sum_{rsq}^{\notin KK'K''}\frac{C^m_{rs}B^m_{sq}}{\omega-E^m_r+E^m_s+i\eta}
\rho_{qr}^{i\to f}(m)e^{i(\omega+E_{s}^{m}-E_{q}^{m}+i\eta)t}\nonumber\\
&+\sum_{m=m_0}^N\sum_{rsr_1s_1}^{\notin KK'K_1K'_1}S^m_{rr_1}\frac{\sum_q\tilde{R}^m_{r_1q}C^m_{qs_1}}{\omega-E^m_{r_1}+E^m_{s_1}+i\eta}S^m_{s_1s}
B^m_{sr}e^{i(\omega+E_{s}^{m}-E_{r}^{m}+i\eta)t}
,\label{eq:gf<-positive-times-t1}
\end{align}

\begin{align}
G^<(\omega,t<0)
=&\sum_{m=m_0}^N\sum_{rs}^{\notin KK'}B^m_{rs}\Big(-\frac{\sum_q \tilde{R}^m_{sq}C^m_{qr}}{\omega+E^m_r-E^m_s+i\eta}
+\frac{\sum_q \tilde{R}^m_{sq}C^m_{qr}}{\omega+E^m_r-E^m_s-i\eta}\Big)\nonumber\\
&-\sum_{m=m_0}^N\sum_{rs}^{\notin KK'}B^m_{rs}\frac{\sum_q \tilde{R}^m_{sq}C^m_{qr}}{\omega+E^m_r-E^m_s-i\eta}
e^{i(\omega+E_{r}^{m}-E_{s}^{m}-i\eta)t}\nonumber\\
&+\sum_{m=m_0}^N\sum_{rsr_1s_1}^{\notin KK'K_1K'_1}\frac{C^m_{rs}}{\omega-E^m_{r}+E^m_{s}-i\eta}S^m_{ss_1}
\sum_q B^m_{s_1q}\tilde{R}^m_{qr_1}e^{i(\omega+E_{s_1}^{m}-E_{r_1}^{m}-i\eta)t}S^m_{r_1r}
.\label{eq:gf<-negative-times-t1}
\end{align}

In the case that $t=t_2$
\begin{align}
G^<(\omega,t>0)
=&\sum_{m=m_0}^N\sum_{rsq}^{\notin KK'K''}\Big(-\frac{C^m_{rs}B^m_{sq}}{\omega-E^m_q+E^m_s+i\eta}
+\frac{C^m_{rs} B^m_{sq}}{\omega-E^m_q+E^m_s-i\eta}\Big)
\rho_{qr}^{i\to f}(m)e^{i(E_{r}^{m}-E_{q}^{m})t}\nonumber\\
&-\sum_{m=m_0}^N\sum_{rsq}^{\notin KK'K''}\frac{C^m_{rs}B^m_{sq}}{\omega-E^m_q+E^m_s-i\eta}
\rho_{qr}^{i\to f}(m)e^{-i(\omega+E_{s}^{m}-E_{r}^{m}-i\eta)t}\nonumber\\
&+\sum_{m=m_0}^N\sum_{rsr_1s_1}^{\notin KK'K_1K'_1}S^m_{rr_1}\frac{\sum_q B^m_{r_1q}\tilde{R}^m_{qs_1}}{\omega+E^m_{r_1}-E^m_{s_1}-i\eta}S^m_{s_1s}
C^m_{sr}e^{-i(\omega-E_{s}^{m}+E_{r}^{m}-i\eta)t}
,\label{eq:gf<-positive-times-t2}
\end{align}

\begin{align}
G^<(\omega,t<0)
=&\sum_{m=m_0}^N\sum_{rs}^{\notin KK'}B^m_{rs}\Big(\frac{\sum_q \tilde{R}^m_{sq}C^m_{qr}}{\omega+E^m_r-E^m_s-i\eta}
-\frac{\sum_q \tilde{R}^m_{sq}C^m_{qr}}{\omega+E^m_r-E^m_s+i\eta}\Big)\nonumber\\
&+\sum_{m=m_0}^N\sum_{rs}^{\notin KK'}B^m_{rs}\frac{\sum_q \tilde{R}^m_{sq}C^m_{qr}}{\omega+E^m_r-E^m_s+i\eta}
e^{-i(\omega+E_{r}^{m}-E_{s}^{m}+i\eta)t}\nonumber\\
&-\sum_{m=m_0}^N\sum_{rsr_1s_1}^{\notin KK'K_1K'_1}\frac{B^m_{rs}}{\omega+E^m_{r}-E^m_{s}+i\eta}S^m_{ss_1}
\sum_q \tilde{R}^m_{s_1q}C^m_{qr_1}e^{-i(\omega-E_{s_1}^{m}+E_{r_1}^{m}+i\eta)t}S^m_{r_1r}
.\label{eq:gf<-negative-times-t2}
\end{align}

\subsection{Greater Green function}

In the case that $t=(t_1+t_2)/{2}$
\begin{align}
&G^>(\omega,t>0)=\nonumber\\
&-\sum_{m=m_0}^N\sum_{rsq}^{\notin KK'K''}B^m_{rs}C^m_{sq}\frac{e^{-i(E^m_q-E^m_r)t}-e^{-2i(\omega+E^m_q-E^m_s)t}e^{-2\eta t}}{\omega-E^m_s+\frac{E^m_q+E^m_r}{2}-i\eta}\rho^{i\to f}_{qr}(m)
+\sum_{m=m_0}^N\sum_{rsq}^{\notin KK'K''}B^m_{rs}C^m_{sq}\frac{e^{-i(E^m_q-E^m_r)t}-e^{2i(\omega+E^m_r-E^m_s)t}e^{-2\eta t}}{\omega-E^m_s+\frac{E^m_q+E^m_r}{2}+i\eta}\rho^{i\to f}_{qr}(m)\nonumber\\
&-\sum_{m=m_0}^N\sum_{rsr_1s_1}^{\notin KK'K_1K_1'}C^m_{rs}e^{-2i(\omega+E^m_r-E^m_s)t}e^{-2\eta t}\frac{S^m_{ss_1}\sum_q \tilde{R}^m_{s_1q}B^m_{qr_1}S^m_{r_1r}}{\omega-\frac{E^m_r-E^m_s+E^m_{r_1}-E^m_{s_1}}{2}-i\eta}
+\sum_{m=m_0}^N\sum_{rsr_1s_1}^{\notin KK'K_1K_1'}B^m_{rs}e^{2i(\omega+E^m_r-E^m_s)t}e^{-2\eta t}\frac{S^m_{ss_1}\sum_qC^m_{s_1q}\tilde{R}^m_{qr_1}S^m_{r_1r}}{\omega+\frac{E^m_r-E^m_s+E^m_{r_1}-E^m_{s_1}}{2}+i\eta},\label{eq:ggreater-tave-positive}
\end{align}

\begin{align}
&G^>(\omega,t<0)=\nonumber\\
&-\sum_{m=m_0}^N\sum_{rs}^{\notin KK'}B^m_{rs}\frac{1-e^{2i(\omega+E^m_r-E^m_s)t}e^{2\eta t}}{\omega+E^m_r-E^m_s-i\eta}\sum_qC^m_{sq}\tilde{R}^m_{qr}
+\sum_{m=m_0}^N\sum_{rs}^{\notin KK'}B^m_{rs}\frac{1-e^{-2i(\omega+E^m_r-E^m_s)t}e^{2\eta t}}{\omega+E^m_r-E^m_s+i\eta}\sum_qC^m_{sq}\tilde{R}^m_{qr}\nonumber\\
&-\sum_{m=m_0}^N\sum_{rsr_1s_1}^{\notin KK'K_1K_1'}C^m_{rs}e^{2i(\omega-E^m_{s_1}+E^m_{r_1})t}e^{2\eta t}\frac{S^m_{ss_1}\sum_q\tilde{R}^m_{s_1q}B^m_{qr_1}S^m_{r_1r}}{\omega-\frac{E^m_r-E^m_s+E^m_{r_1}-E^m_{s_1}}{2}-i\eta}
+\sum_{m=m_0}^N\sum_{rsr_1s_1}^{\notin KK'K_1K_1'}B^m_{rs}e^{-2i(\omega+E^m_{s_1}-E^m_{r_1})t}e^{2\eta t}\frac{S^m_{ss_1}\sum_qC^m_{s_1q}\tilde{R}^m_{qr_1}S^m_{r_1r}}{\omega+\frac{E^m_r-E^m_s+E^m_{r_1}-E^m_{s_1}}{2}+i\eta}\label{eq:glesser-t-negative}.
\end{align}

In the case that $t=t_1$
\begin{align}
G^>(\omega,t>0)
=&\sum_{m=m_0}^N\sum_{rsq}^{\notin KK'K''}\Big(-\frac{B^m_{rs}C^m_{sq}}{\omega-E^m_s+E^m_q-i\eta}
+\frac{B^m_{rs} C^m_{sq}}{\omega-E^m_s+E^m_q+i\eta}\Big)
\rho_{qr}^{i\to f}(m)e^{i(E_{r}^{m}-E_{q}^{m})t}\nonumber\\
&-\sum_{m=m_0}^N\sum_{rsq}^{\notin KK'K''}\frac{B^m_{rs}C^m_{sq}}{\omega-E^m_s+E^m_q+i\eta}
\rho_{qr}^{i\to f}(m)e^{i(\omega-E_{s}^{m}+E_{r}^{m}+i\eta)t}\nonumber\\
&+\sum_{m=m_0}^N\sum_{rsr_1s_1}^{\notin KK'K_1K'_1}S^m_{rr_1}\frac{\sum_q{C}^m_{r_1q}\tilde{R}^m_{qs_1}}{\omega-E^m_{r_1}+E^m_{s_1}+i\eta}S^m_{s_1s}
B^m_{sr}e^{i(\omega+E_{s}^{m}-E_{r}^{m}+i\eta)t}
,\label{eq:gf>-positive-times-t1}
\end{align}

\begin{align}
G^>(\omega,t<0)
=&\sum_{m=m_0}^N\sum_{rs}^{\notin KK'}B^m_{rs}\Big(\frac{\sum_q C^m_{sq}\tilde{R}^m_{qr}}{\omega-E^m_s+E^m_r+i\eta}
-\frac{\sum_q C^m_{sq}\tilde{R}^m_{qr}}{\omega-E^m_s+E^m_r-i\eta}\Big)\nonumber\\
&+\sum_{m=m_0}^N\sum_{rs}^{\notin KK'}B^m_{rs}\frac{\sum_q C^m_{sq}\tilde{R}^m_{qr}}{\omega-E^m_s+E^m_r-i\eta}
e^{i(\omega+E_{r}^{m}-E_{s}^{m}-i\eta)t}\nonumber\\
&-\sum_{m=m_0}^N\sum_{rsr_1s_1}^{\notin KK'K_1K'_1}\frac{C^m_{rs}}{\omega-E^m_{r}+E^m_{s}-i\eta}S^m_{ss_1}
\sum_q \tilde{R}^m_{s_1q}B^m_{qr_1}e^{i(\omega+E_{s_1}^{m}-E_{r_1}^{m}-i\eta)t}S^m_{r_1r}
.\label{eq:gf>-negative-times-t1}
\end{align}

In the case that $t=t_2$
\begin{align}
G^>(\omega,t>0)
=&\sum_{m=m_0}^N\sum_{rsq}^{\notin KK'K''}\Big(\frac{B^m_{rs}C^m_{sq}}{\omega-E^m_s+E^m_r+i\eta}
-\frac{B^m_{rs} C^m_{sq}}{\omega-E^m_s+E^m_r-i\eta}\Big)
\rho_{qr}^{i\to f}(m)e^{i(E_{r}^{m}-E_{q}^{m})t}\nonumber\\
&+\sum_{m=m_0}^N\sum_{rsq}^{\notin KK'K''}\frac{B^m_{rs}C^m_{sq}}{\omega-E^m_s+E^m_r-i\eta}
\rho_{qr}^{i\to f}(m)e^{-i(\omega+E_{q}^{m}-E_{s}^{m}-i\eta)t}\nonumber\\
&-\sum_{m=m_0}^N\sum_{rsr_1s_1}^{\notin KK'K_1K'_1}S^m_{rr_1}\frac{\sum_q \tilde{R}^m_{r_1q}B^m_{qs_1}}{\omega+E^m_{r_1}-E^m_{s_1}-i\eta}S^m_{s_1s}
C^m_{sr}e^{-i(\omega-E_{s}^{m}+E_{r}^{m}-i\eta)t}
,\label{eq:gf>-positive-times-t2}
\end{align}

\begin{align}
G^>(\omega,t<0)
=&\sum_{m=m_0}^N\sum_{rs}^{\notin KK'}B^m_{rs}\Big(-\frac{\sum_q C^m_{sq}\tilde{R}^m_{qr}}{\omega-E^m_s+E^m_r-i\eta}
+\frac{\sum_q C^m_{sq}\tilde{R}^m_{qr}}{\omega-E^m_s+E^m_r+i\eta}\Big)\nonumber\\
&-\sum_{m=m_0}^N\sum_{rs}^{\notin KK'}B^m_{rs}\frac{\sum_q C^m_{sq}\tilde{R}^m_{qr}}{\omega-E^m_s+E^m_r+i\eta}
e^{-i(\omega+E_{r}^{m}-E_{s}^{m}+i\eta)t}\nonumber\\
&+\sum_{m=m_0}^N\sum_{rsr_1s_1}^{\notin KK'K_1K'_1}\frac{B^m_{rs}}{\omega+E^m_{r}-E^m_{s}+i\eta}S^m_{ss_1}
\sum_q C^m_{s_1q}\tilde{R}^m_{qr_1}e^{-i(\omega-E_{s_1}^{m}+E_{r_1}^{m}+i\eta)t}S^m_{r_1r}
.\label{eq:gf>-negative-times-t2}
\end{align}

\end{widetext}
\section{Convergence of the Lorentzian broadening scheme for
  time-dependent spectral functions}
\label{sec:appendix-broadening}
Within the NRG approach, equilibrium Green functions have a discrete
Lehmann representation consisting a set of poles at the excitations of
the system. Replacing the delta functions in the imaginary part of the Green functions with
Gaussian or logarithmic-Gaussians \cite{Sakai1989,Bulla2001,Bulla2008}
yields smooth spectral functions $A(\omega)$. For nonequilibrium Green
functions, and their associated time-dependent spectral functions $A(\omega,t)$, we
argued in Sec.~\ref{sec:gf-times}, that a Lorentzian broadening procedure
is required to consistently broaden the regular and singular parts
contributing to the imaginary part of the nonequilibrium Green
function. Since Lorentzians have long tails, compared to the
exponential ones of Gaussians, it is important to check the
convergence w.r.t. to the value of the broadening parameter used,
which we do here. Another issue which arose in Sec.~\ref{sec:glesser},
concerned the origin of the positive spectral weight in time-dependent occupied density of states
$\pi\Gamma N(\omega,t)=\Gamma G^{<}(\omega,t)/(2i)$ which
is found even at $T=0$ and in the long time limit $t\to +\infty$. In
particular, whether this might be attributed to the use of a
Lorentzian broadening scheme. We show that this is not the
case. Instead, as discussed in more detail in
Appendix~\ref{sec:appendix-thermalization}, it is a result of
imperfect thermalization within the TDNRG approach.

\begin{figure}[t]
    \centering 
  	\includegraphics[width=0.482\textwidth]{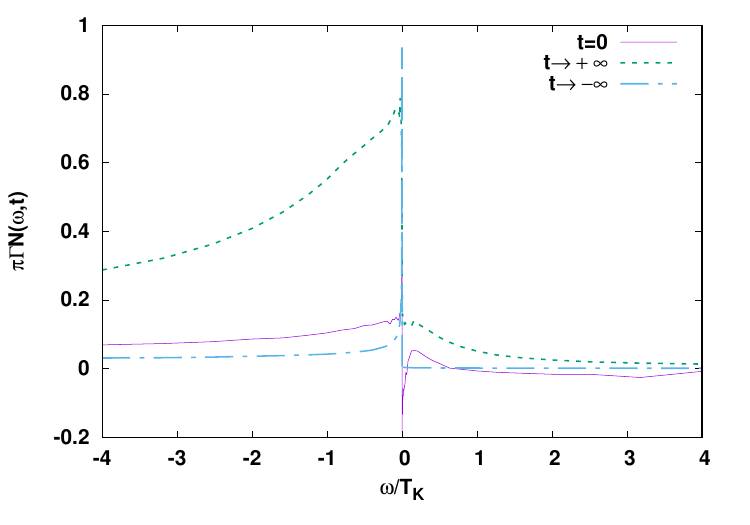}
    \caption{The normalized occupied density of states
      $\pi\Gamma N(\omega,t)=\Gamma G^{<}(\omega,t)/(2 i)$ vs $\omega$ as in
      Fig.~\ref{fig:fig6}
      at three different average times $t=-\infty,0,+\infty$, with
      Lorentzian broadening parameter $\eta_0=1/N_z=1/32$
      and in the frequency range close to the Fermi level.}
    \label{fig:fig9}
\end{figure}

We refer to Fig.~\ref{fig:fig6} showing the time-dependent occupied density of
states $N(\omega,t)$ defined from the lesser Green function and evaluated
by using the Lorentzian broadening. 
One may see that the density of states is finite even at positive
frequency and long times even though the temperature is zero.
This is different from the equilibrium lesser Green function at zero
temperature which only gives a finite density of state below the Fermi
level ($\omega =0$) as follows from the equilibrium result in
Eq.~(\ref{eq:fd1}). It is not obvious that the non-zero density in
$N(\omega>0,t\to +\infty)$  is due to the broadening scheme or due to the nonequilibrium effect or both. 
Figure~\ref{fig:fig9} shows $N(\omega,t)$ at three different times;
infinite past, zero time, and infinite future but in the frequency range closer to the Fermi energy level.
It is clear that the occupied density of states in the infinite past should be
equal to the occupied density of states in the equilibrium initial state, which
by Eq.~(\ref{eq:fd1}) implies a zero occupied density of states for $\omega>0$,
as indeed observed. In contrast, at zero time, the occupied density of states shows both positive and negative values at positive frequencies.
and in the infinite future, the occupied density of states shows a finite
positive value at $\omega>0$. This figure already suggests that
imperfect thermalization at long positive times leads to the non-zero
occupied density of states for $\omega>0$.

\begin{figure}[t]
    \centering 
  	\includegraphics[width=0.482\textwidth]{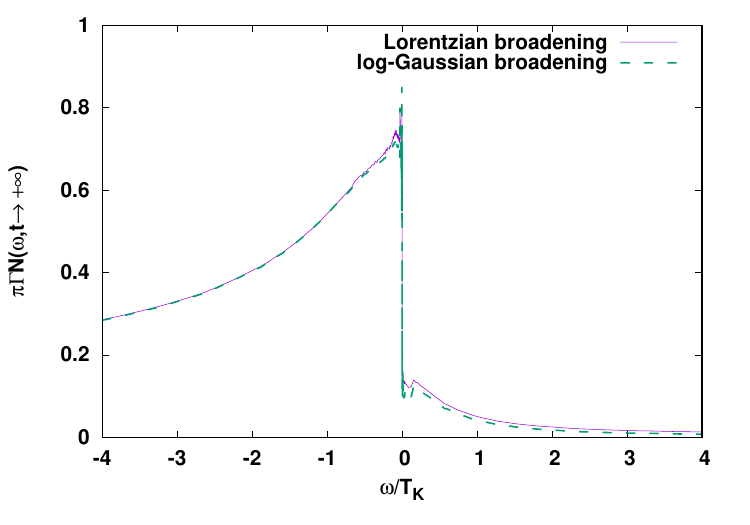}
    \caption{The normalized occupied density of states $\pi\Gamma N(\omega,t)=
      \Gamma G^{<}(\omega,t)/(2 i)$ vs $\omega/T_{\rm K}$ as in
      Fig.~\ref{fig:fig6} at average-time $t=+\infty$ and in the
      frequency range close to the Fermi level,
      calculated with the Lorentzian and logarithmic-Gaussian broadening.}
    \label{fig:fig10}
\end{figure}

To shed light on the above problem, we also calculate the zero
temperature occupied density of
state $N(\omega,t)$ in the infinite future using a logarithmic-Gaussian broadening.
This is possible since for $t\to +\infty$ only the pole contributions
to the lesser Green function remain, and the expression 
can be reduced to a set of delta functions, for which the usual
logarithmic-Gaussian broadening applies.
Figure \ref{fig:fig10} shows the comparison of $N(\omega,t)$ determined
with the two different broadening schemes, Lorentzian and
logarithmic-Gaussian and using the same value of $\eta_0=1/N_z=1/32$
where $\eta_0$ is related to the infinitesimal broadening $\eta$
appearing in the Green functions by $\eta = \eta_0 |\Delta E|$, with
$\Delta E$ an excitation appearing in the Green function.
One sees that both schemes give nearly identical results, and
moreover, both schemes result in positive spectral weight at
$\omega>0$.

\begin{figure}[t]
    \centering 
  	\includegraphics[width=0.482\textwidth]{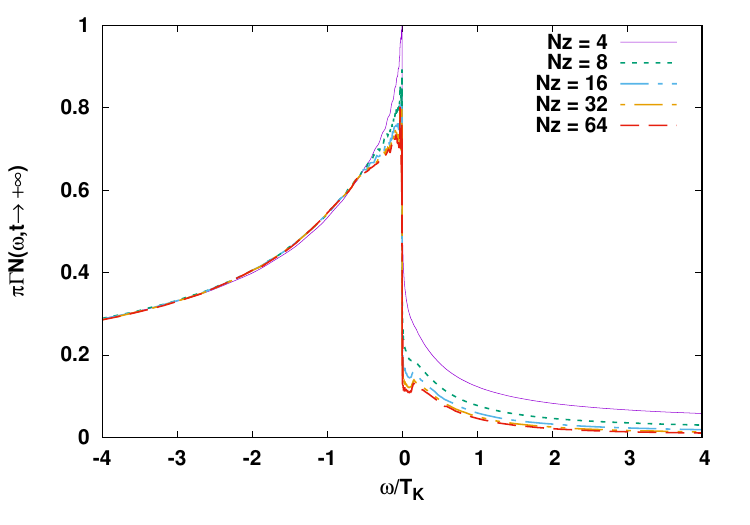}
    \caption{The normalized occupied density of states $\pi\Gamma N(\omega,t)=
      \Gamma G^{<}(\omega,t)/(2 i)$ vs $\omega/T_{\rm K}$ as in Fig.~\ref{fig:fig6} at average-time $t=+\infty$ and in the frequency range close to the Fermi level, calculated with the Lorentzian broadening and $\eta_0=1/N_z$.}
    \label{fig:fig11}
\end{figure}

\begin{figure}[t]
    \centering 
  	\includegraphics[width=0.482\textwidth]{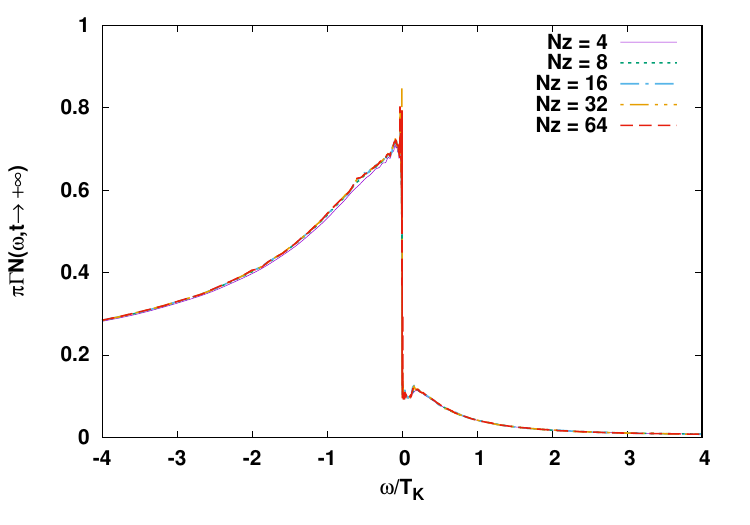}
    \caption{The normalized occupied density of states $\pi\Gamma N(\omega,t)=
      \Gamma G^{<}(\omega,t)/(2 i)$ vs $\omega/T_{\rm K}$ as in
      Fig.~\ref{fig:fig6} at average time $t=+\infty$ and in the
      frequency range close to the Fermi level,
      calculated with the logarithmic-Gaussian broadening and $\eta_0=1/N_z$.}
    \label{fig:fig12}
\end{figure}

It is well known that the function $1/(\omega-\omega_0+i\eta)$, within the Lorentzian broadening scheme, 
decays slowly away from $\omega_0$, while the same function with the
same value of $\eta_0$ approximated by the logarithmic-Gaussian is more
local. Therefore, for the Lorentzian broadening, the smaller the $\eta_0$
the more accurate the result. In contrast, for the logarithmic-Gaussian
broadening, the result is less sensitive to the precise value of $\eta_0$.
This is illustrated in Figs.~\ref{fig:fig11} and \ref{fig:fig12}, which show $N(\omega,t)$ at the infinite future
using the Lorentzian and logarithmic-Gaussian broadening schemes, respectively, and for different values of $\eta_0=1/N_z$.

In Fig.~\ref{fig:fig11}, the results with the Lorentzian broadening
shows a strong dependence on the value of $\eta_0$.
The results starts to converge when $\eta_0$ is as small as $1/32$. In
contrast, the results with the logarithmic-Gaussian broadening in Fig.~\ref{fig:fig12} shows a much weaker dependence on $\eta_0$.
We conclude that the Lorentzian broadening
scheme yields converged results for spectral functions for $\eta_0=1/32$ and
that the observed finite spectral weight in $N(\omega,t\to +\infty$ at
$\omega>0$ is not an artefact of the Lorentzian broadening as the
same result is found for the logarithmic-Gaussian scheme.

\section{Thermalization}
\label{sec:appendix-thermalization}

\begin{figure}[t]
    \centering 
  	\includegraphics[width=0.482\textwidth]{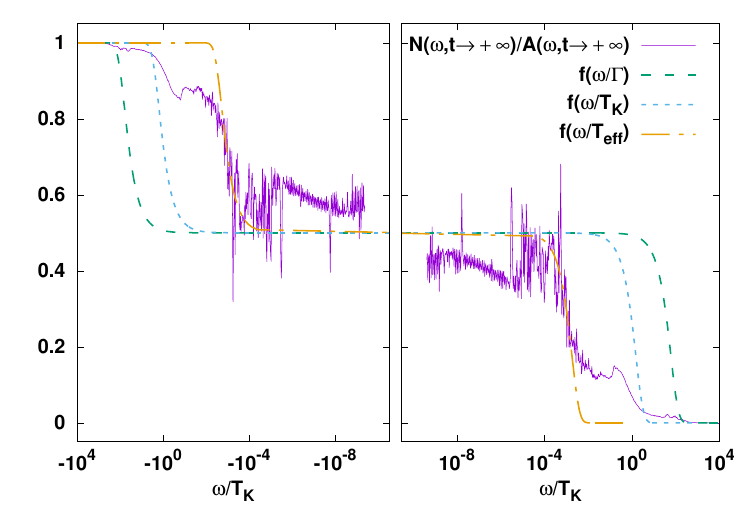}
    \caption{Effective Fermi distribution $f_{\rm
        eff}(\omega)=N(\omega,t\to +\infty)/A(\omega,t\to +\infty)$ vs
      $\omega/T_{\rm K}$
      where $\varepsilon^i_d=-U^i/2=-15\Gamma$ and $T^i_{\rm K}=3\times 10^{-8}D=3\times 10^{-5}\Gamma$.}
    \label{fig:fig13}
\end{figure}

\begin{figure}[t]
    \centering 
  	\includegraphics[width=0.482\textwidth]{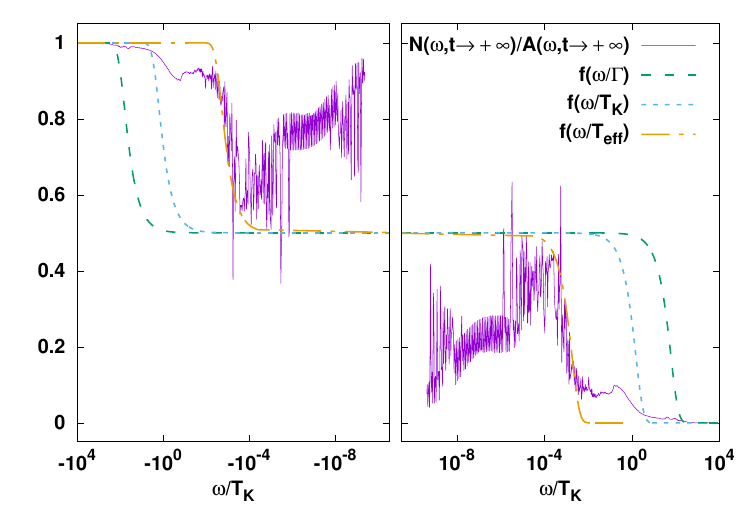}
    \caption{Effective Fermi distribution $f_{\rm
        eff}(\omega)=N(\omega,t\to +\infty)/A(\omega,t\to +\infty)$  vs
      $\omega/T_{\rm K}$
      where $\varepsilon^i_d=-U^i/2=-12\Gamma$ and $T^i_{\rm K}=3\times 10^{-7}D=3\times10^{-4}\Gamma$.}
    \label{fig:fig14}
\end{figure}

\begin{figure}[t]
    \centering 
  	\includegraphics[width=0.482\textwidth]{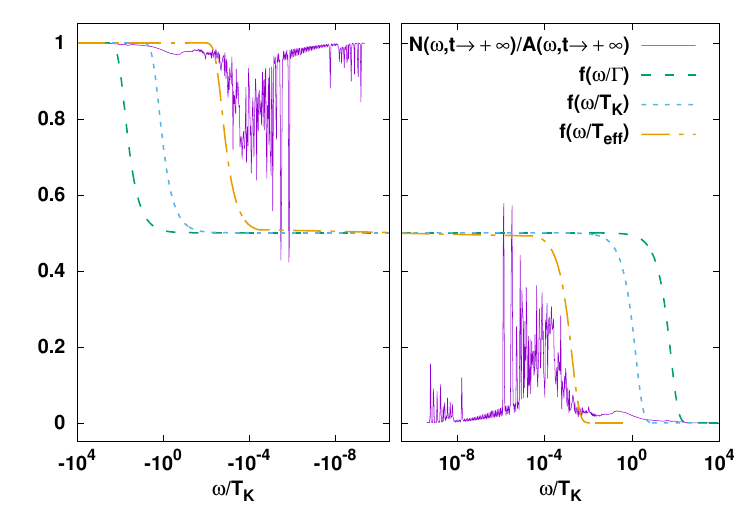}
    \caption{Effective Fermi distribution $f_{\rm
        eff}(\omega)=N(\omega,t\to +\infty)/A(\omega,t\to +\infty)$  vs
      $\omega/T_{\rm K}$
      where $\varepsilon^i_d=-U^i/2=-9\Gamma$ and $T^i_{\rm K}=2.8\times 10^{-6}D=2.8\times 10^{-3}\Gamma$.}
    \label{fig:fig15}
\end{figure}

The observation of a non-zero occupied density of states at positive
frequency at the infinite future and for zero temperature indicates
imperfect thermalization of the system in this limit. 
This is due to the use of a discrete conduction electron bath in the
NRG approach, which in nonequilibrium situations cannot properly
dissipate the energy change following a sudden quench due to the
nonextensive heat capacity of the discrete Wilson chain bath \cite{Rosch2012,Nghiem2014a,Nghiem2018}
We expect that for a true heat bath, that the occupied density of
states at infinite time will follow the expression (\ref{eq:fd1}) in
the main text. To investigate the problem in more detail, we calculate the "effective" Fermi distribution 
which is defined by $N(\omega,t)/A(\omega,t)$
when $t$ is in the infinite time limit, which we denote as $f_{\rm eff}(\omega)$. 
The results are shown in Figs.~\ref{fig:fig13}-\ref{fig:fig15}, 
in which the calculations are done with the same final state $\varepsilon^f_d=-U^f/2=-6\Gamma$ and three different 
initial states $\varepsilon^i_d=-U^i/2=-15\Gamma$, $-12\Gamma$, and $-9\Gamma$, where $\Gamma=0.001$. 

We see that the effective Fermi distribution does not follow the Fermi
distribution at the effective temperature $T_{\rm eff}=\Gamma$ or $T_{\rm K}$, 
but only shows deviations from the Fermi distribution at these temperature.
However, all three effective Fermi distributions follow the Fermi
distribution with an effective temperature $T_{\rm eff}=3\times
10^{-8}$ at low frequencies (only by coincidence, this is close to the initial state Kondo temperature
of one of the three quenches in Figs.~\ref{fig:fig13}-\ref{fig:fig15},
namely that in Fig.~\ref{fig:fig13}).
Therefore, we conclude that the long-time limit is independent of the
initial state, but that some heating up occurs in the evolution towards the
final state leading to an imperfect thermalization at $t=+\infty$.
The amount of this heating up is relatively small since
$T_{\rm eff}/T_{\rm K}=1.2\times 10^{-3}\approx \Gamma/D$.

\bibliography{noneq-nrg}
\end{document}